\documentclass[10pt,english,aps,prb,reprint,showpacs]{revtex4-1}
\usepackage{amsmath}
\usepackage{graphicx}

\usepackage{bm}
\usepackage{babel}
\usepackage{multirow}
\usepackage{setspace}

\makeatother
\begin{document}

\title{Identification of Structure in Condensed Matter with the Topological Cluster Classification}
\author{Alex Malins}
\affiliation{Bristol Centre for Complexity Sciences, University of Bristol, Bristol, BS8 1TS, UK}
\affiliation{School of Chemistry, University of Bristol, Cantock's Close, Bristol, BS8 1TS, UK}
\author{Stephen R. Williams}
\affiliation{Research School of Chemistry, The Australian National University, Canberra, ACT 0200, Australia}
\author{Jens Eggers}
\affiliation{School of Mathematics, University of Bristol, University Walk, Bristol, BS8 1TW, UK}
\author{C. Patrick Royall}
\affiliation{HH Wills Physics Laboratory, Tyndall Avenue, Bristol, BS8 1TL, UK}
\affiliation{School of Chemistry, University of Bristol, Cantock's Close, Bristol, BS8 1TS, UK}
\affiliation{Centre for Nanoscience and Quantum Information, Tyndall Avenue, Bristol, BS8 1FD, UK}

\email{paddy.royall@bristol.ac.uk}

\date{\today}

\begin{abstract}
We describe the topological cluster classification (TCC) algorithm. The TCC detects local structures with bond topologies similar to isolated clusters which minimise the potential energy for a number of monatomic and binary simple liquids with $m\leq13$ particles. We detail a modified Voronoi bond detection method that optimizes the cluster detection. The method to identify each cluster is outlined, and a test example of Lennard-Jones liquid and crystal phases is considered and critically examined.
\end{abstract}

\maketitle

\section{Introduction}

Research into the structure of disordered systems, such as liquids, glasses, liquid-vapor interfaces and cluster fluids, has a rich history. Early studies focused on pair distribution functions as these were readily accessible from experiments using scattering techniques. The experimental results inspired the development of theories with the capability to predict the radial distribution functions for disordered states. The most famous example of such a theory is perhaps the Percus-Yevick closure of the Ornstein-Zernike relation which provides the radial distribution function for the hard sphere fluid \cite{hansen}.

Efforts to understand the structure of liquids in greater detail than at the level of pair correlations of the particle density were pioneered by Bernal, Rahman and Finney \cite{bernal1959,rahman1966,finney1970a}. These authors studied the properties of Voronoi cells in liquids, which are convex polyhedra that contain all points in space closer to one particle than any other. The geometrical characteristics of each polyhedron, such as the number and shape of the faces, depends on the position of the particle and its immediate neighbors. As such, Voronoi cells provide information on higher-degree correlations of the particle density than two-body distribution functions. 

These studies into Voronoi polyhedra were among some of the earliest that sought to characterize the structure of disordered systems in terms of the shapes formed by small clusters of particles. Many subsequent efforts have been made in this direction, and there are now several methods available for quantifying structural correlations in disordered systems in terms of the shapes of clusters of particles. We designate methods as \textit{topological} methods if they identify clusters from topological features of the ``bond network'' formed by the particle locations and the connections between neighboring particles.

Two topological methods that see frequent use are the common neighbor analysis (CNA) introduced by Honeycutt and Andersen \cite{honeycutt1987} and the Voronoi face analysis (VFA) by Tanemura \textit{et al.}\ \cite{tanemura1977}. The CNA identifies the structural ordering around pairs of particles in terms of shared (common) neighboring particles. The VFA technique is a generalisation of the early Voronoi cell studies. The structure around a single particle is characterized by the arrangement of its nearest neighbors and the bonds between them. This information is encoded by the shapes of the faces of its Voronoi cell. 
Both of these methods have been used to study phenomena including the melting and freezing of clusters \cite{honeycutt1987}, structure in supercooled liquids on cooling towards the glass transition \cite{jonsson1988,tsumuraya1990,dellavalle1994,bailey2004,coslovich2007a,hedges2009}, and crystallisation \cite{tanemura1977,hsu1979a,swope1990}.

An alternative approach is the bond orientational order parameters, introduced by Steinhardt \emph{et al.}\ to characterize the arrangement of neighboring particles around a central particle \cite{steinhardt1981,steinhardt1983}. A series of bond orientation order (BOO) parameters are defined that quantify the similarity between the directions from a particle to its neighbors and the spherical harmonic solutions to the Laplace equation, in a method that has been described as a kind of \textit{shape spectroscopy} \cite{steinhardt1983,mitus1996}.
BOO parameters have been applied for the study of a variety of phenomena. Their original application was to study structural ordering in supercooled liquids \cite{steinhardt1981,steinhardt1983}, where depending upon the system under consideration, both icosahedral- \cite{tomida1995} and crystalline-ordering \cite{tanaka2010} has been found to develop on cooling towards the glass transition. Other researchers have employed BOO parameters to study the structure at interfaces \cite{heni2002,wierschem2011} and of large clusters \cite{chakravarty2002}. Arguably the most successful application of BOO parameters however is as order parameters for crystallisation \cite{tenwolde1995,auer2004,lechner2008,vanmeel2009,kawasaki2010b,russo2012}.

Recent advances in experimental techniques on colloidal dispersions have delivered methods that allow measurement of real space configurational data for colloids. In particular, particle tracking software allows the positions to be extracted from images taken with 3D confocal microscopy \cite{vanblaaderen1995,crocker1995,royall2003}. This development means that the structure of experimental systems that display Brownian motion can now be analysed with higher-order structural detection algorithms, with interest in, for example, colloidal gels \cite{royall2008}, glasses \cite{vanblaaderen1995} and critical fluctuations \cite{royall2007c}.

Here we describe a recent addition to the collection of high-order structural detection algorithms, the topological cluster classification (TCC) algorithm. This algorithm has been used to investigate higher-order structure in colloidal \cite{royall2008,royall2008aip} and molecular \cite{royall2011c60} gels, colloid-polymer mixtures \cite{taffs2010jcp}, colloidal clusters \cite{malins2009,malins2010,klix2013},
simple liquids \cite{taffs2010}, liquid-vapor interfaces \cite{godonoga2010}, supercooled liquids \cite{malins2013jcp,malins2013fara,royall2014} and crystallising fluids \cite{taffs2013}.
The method shares some characteristics with other topological methods (e.g.\ the VFA and CNA methods). However, rather than detecting clusters around a central or pair of particles, it is based on the detection of minimum energy clusters with a range of sizes. By minimum energy cluster, we refer to those structures which minimize the potential energy of a given number of particles in isolation. Here we consider spherically symmetric interactions, whose minimum energy configurations can be calculated by optimization methods such as in the GMIN software \cite{wales1997}.

The philosophy of the TCC algorithm draws heavily upon an idea proposed by Sir Charles Frank in a seminal paper in 1952 \cite{frank1952}. When discussing the ability to stabilize metallic liquids below the melting temperature, Frank proposed that collections of particles may preferentially adopt certain energy-minimising structures on short lengthscales. He inferred that if the symmetry of the structures formed is incommensurate with that of the bulk crystal symmetry, the structures would inhibit crystallisation and stabilize the undercooled melt. The basis for this was that there must be a substantial rearrangement of the structures for crystallisation to proceed, and this operation would carry an energetic cost.

To demonstrate his idea Frank cited the example of the minimum energy clusters formed by $13$ Lennard-Jones particles in isolation. Arranging the particles at the centre and the vertices of a regular icosahedron yields a cluster with $8.4\%$ lower interaction energy than for compact FCC or HCP crystal clusters of 13 particles. Assuming that the energy of the icosahedral arrangement would still be lowest for the three arrangements in the disordered liquid environment, he inferred that the icosahedral formations would be more prevalent than FCC and HCP arrangements in the supercooled liquid. This turned out to be a reasonable assumption, the icosahedron was shown to have lower energy than either crystal arrangement in a mean-field description of the Lennard-Jones liquid \cite{mossa2003}.

While Frank considered minimum energy clusters of 13 Lennard-Jones particles for his example, as a natural extension, the TCC algorithm takes the minimum energy clusters of other numbers of particles as well. The algorithm also considers models with interactions other than the Lennard-Jones potential. The TCC algorithm works by searching a configuration for arrangements of $m$ particles whose bond network is similar to that found in the minimum energy clusters of a given model. By identifying minimum energy clusters of the model of interest, the method therefore provides a direct link between the interactions in the system under consideration and any structural ordering that is found. Moreover, the TCC simultaneously finds clusters for all models incorporated. Thus an analysis of a given coordinate set is not limited to the minimum energy clusters which correspond to the model with which the set was generated.

The TCC favors no specific ``origin'' particles for the clusters, i.e.\ a central particle or a bonded pair. This means that routines can be devised to detect clusters with disparate shapes and sizes where there are no such ``origin'' particles. This stance differentiates the TCC from many of the other structure detection methods. The algorithm presented here includes detection routines for minimum energy clusters of up to $13$ particles for the models that are considered. Extending the algorithm to include detection routines for larger minimum energy clusters is possible.
This paper is organized as follows. In section \ref{sectionOverview}, we give a high-level overview of the method, and proceed, section  \ref{sectionVoronoi}, to detail the Voronoi construction devised to optimize the structure detection algorithm. In section \ref{sectionTCC} and the appendix, we provide details of the TCC detection routines for specific clusters. In section \ref{sectionTCCexample}, we demonstrate the TCC algorithm by identifying the structure in Lennard-Jones crystals. In section \ref{sectionResults} we critically analyse its performance on four phases of the Lennard-Jones system before concluding in section \ref{sectionSummary}.

\section{Overview of the Topological Cluster Classification}
\label{sectionOverview}

A high-level overview of the TCC algorithm is as follows:

\begin{enumerate}
\item The neighbors of each particle are identified.
\item The network formed by the particles and connections to their neighbors is searched for shortest-path rings of $3$, $4$ and $5$ particles.
\item From the shortest-path rings, a set of structures known as the ``basic clusters'' are identified. The basic clusters are distinguished by the number of additional particles that are common neighbors of all the particles in a shortest-path ring.
\item Larger clusters are then identified by combining basic clusters together, sometimes with the addition of one or two separately bonded particles, according to a set of predefined rules. The method yields structures with bond networks with similar topology to the bond networks of the minimum energy clusters.
\end{enumerate}

Throughout we endeavor to highlight the strengths and weaknesses of the TCC method. Although the focus of the TCC is on disordered systems, the performance of different structure detection algorithms for identifying crystalline order will provide a helpful benchmark for the accuracy and efficiency of the algorithm. These tests are useful because the ordering of particles in a crystal is known \textit{a priori}, whereas this is not necessarily the case for a disordered system.

\section{Detecting neighbors with Voronoi tessellations}
\label{sectionVoronoi}

Although simple and intuitive, the main disadvantage of using a cut-off length to determine a particle's neighbors is the sensitivity of the result to the chosen length and the consequential difficulty in deciding which length yields the most valid result. A method that can identify the neighbor network independently of an adjustable length parameter is therefore appealing. One such method is the Voronoi tessellation which decomposes space into regions of non-intersecting domains with distinct boundaries \cite{voronoi1908}. Each region contains the position of a single particle and all points in space closer to that particle than any other. In three dimensions these regions are convex polyhedra that are termed Voronoi cells. Two particles are neighbors if their Voronoi cells share a face.
Two particles with Voronoi cells that share only an edge or a vertex are not considered bonded. It is important to make this clear as later on we will consider the detection of four-membered rings of particles with the Voronoi method. In practice fluctuations of the particle positions in thermal systems mean that edge- and vertex-sharing polyhedra are rare.

The properties of Voronoi cells for disordered systems were first studied 
by Bernal \cite{bernal1959}, Rahman \cite{rahman1966} and Finney \cite{finney1970a}.
Subsequently the method of using a Voronoi tessellation to define neighbors in a network was popularized in studies of crystallisation \cite{tanemura1977} and glass-formers \cite{nelson1983}.
The early studies of crystallisation noted that the detection of the crystal-order when using the Voronoi method was highly sensitive to thermal fluctuations of the particle positions \cite{tanemura1977,hsu1979,hsu1979a}. The Voronoi cell of a particle in a $T=0$ FCC crystal contains six vertices that are each common to four faces. Small thermal vibrations lead to these vertices being split into two vertices joined by a new edge. This adds a new common face between two of the Voronoi cells of the neighboring particles, and consequently adds a new bond to the system. The additional bond complicates the detection of the crystalline order, as there are a wide variety of Voronoi polyhedra in the thermal system corresponding to a single crystal structure.

This problem is equivalent to the unwanted bond forming across a square of particles as discussed in section \ref{sectionModifiedVoronoi} below. Poor performance is often found when trying to detect structures with four-fold symmetries (e.g.\ octahedral structures) when employing a structure detection method upon the Voronoi neighbor network. For this reason numerous authors have proposed modifications to the Voronoi method in order that the detected bond network is more robust against change by thermal fluctuations of the particles \cite{hsu1979a,galashev1980,medvedev1985,swope1990,omalley2001}. 
The sensitivity of the Voronoi method to thermal fluctuations can be demonstrated by examining the number of first-nearest neighbors it detects for FCC and HCP crystals. Frequently second-nearest neighbors are misdetected as first-nearest neighbors \cite{swope1990,omalley2001,omalley2003}, meaning that the number of the first-nearest neighbors is inflated
 \cite{troadec1998}.

In the case of the TCC, one of us (SRW), proposed a modification for the Voronoi method when introducing the TCC structure algorithm to address this issue \cite{williams2007}. The method adds a dimensionless parameter that sets the maximum amount of distortion that a four-membered ring in a plane can undergo before a bond forms between particles at opposite vertices. Good performance was found for detection of octahedral structures in the hard sphere liquid \cite{williams2007}, and here also we demonstrate that detection of FCC and HCP order is improved using the modified Voronoi bond detection method.

The Voronoi method is attractive in that the neighbor network can be identified without having to select an adjustable parameter. It provides a natural and intuitive way to determine the neighbors for each particle. The method has problems when thermal vibrations destroy degenerate vertices and this scenario is especially relevant for the detection of crystalline order if the number of first-nearest neighbors for each particle is over-estimated. 

\subsection{A modified Voronoi method}
\label{sectionModifiedVoronoi}

\begin{figure*}
\begin{centering}
\includegraphics[width=14 cm]{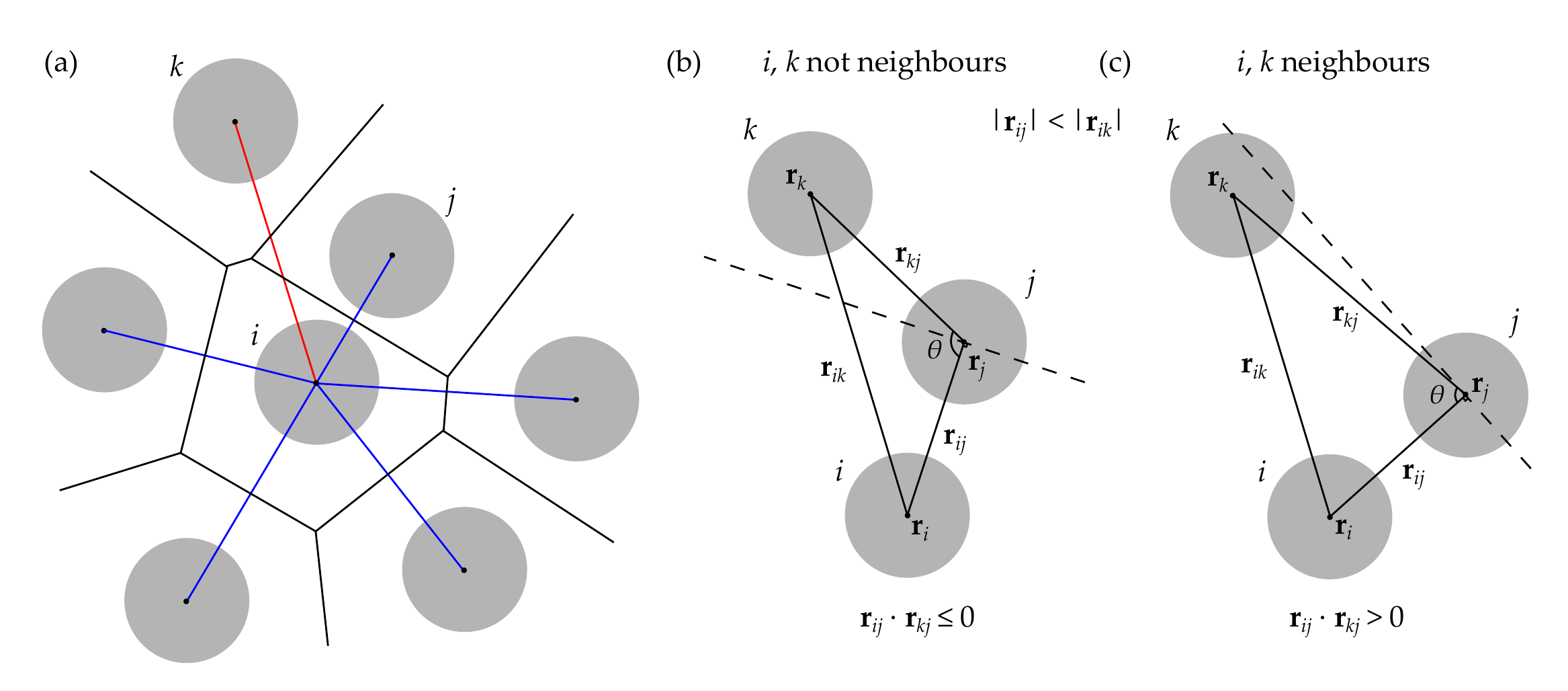} 
\par\end{centering}
\caption{(a) The lines that connect particle centres must also intersect the face that is shared between the two Voronoi cells for the particles to be bonded. Particles $i$ and $k$ are not bonded therefore, as highlighted by the red line connecting their positions. Conversely all other particles are considered neighbors of particle $i$ (blue lines - direct Voronoi neighbors). Particle $j$ is closer to $i$ than $k$, and the volume of its Voronoi cell is shielding particle $k$ from particle $i$. (b) and (c) Identification of direct Voronoi neighbors. Particle $k$ is further from $i$ than $j$ in both cases. In (b) particle $k$ is not bonded to $i$ as there exists $j$ that is closer to $i$ and the angle subtended by vectors $\mathbf{r}_{ij}$ and $\mathbf{r}_{kj}$ is greater than $\pi / 2$ radians. Conversely $k$ is bonded to $i$ in (c) as the bond there is no shielding by $j$.}
\label{figCrossFace} 
\end{figure*}

Our modified Voronoi method makes two modifications to the original Voronoi method (hereafter referred to as the \textit{standard} Voronoi method). The modifications are introduced to improve the detection of the four- and five-membered rings of particles \cite{williams2007}.
The first modification ensures that particles are neighbors only if their Voronoi cells share a face \textit{and} the line that connects the particle positions intersects the shared face. Pairs that obey this condition are known as ``direct Voronoi neighbors'' \cite{meijering1953,brostow1978,medvedev1986}. Conversely neighbors within the standard Voronoi method for which this condition is not obeyed are termed ``indirect Voronoi neighbors''. This condition is depicted in Fig.\ \ref{figCrossFace}(a) for a 2D system, where the connections between direct Voronoi neighbors are highlighted with blue lines, and with red line for the indirect Voronoi neighbors. The second modification is to introduce a dimensionless parameter $f_\mathrm{c}$ that controls the maximum degree of asymmetry of a four-fold ring of neighbors before it is detected instead as two three-fold rings of neighbors.

The first condition has two effects when compared to the standard Voronoi method: 

\begin{enumerate}
\item It strictly reduces the number of bonds derived from the Voronoi tessellation. The removed bonds occur when the volume of the Voronoi cell of a third particle overlaps with the line connecting the positions of two particles that share a Voronoi face. The third particle shields the bond between the other pair and the bond present in the standard Voronoi method is removed [Fig.\ \ref{figCrossFace}(a)]. The removed bonds tend to be ``long bonds'' that cause misdetection of second-nearest neighbors as first-nearest neighbors. 
\item Algorithms to detect neighbors that enforce this condition are generally more straightforward to implement than algorithms which detect all neighbors of the standard Voronoi method.
\end{enumerate}


The direct Voronoi neighbors of particle $i$ are identified by first finding all particles within some cut-off distance $r_{\mathrm{c}}$, which can be made as large as necessary. These particles are ordered in increasing distance from $i$. A particle $k$ is a direct Voronoi neighbor of $i$ if and only if for all $j$ closer to $i$ than $k$ is the angle subtended by the vectors $\mathbf{r}_{ij}$ and $\mathbf{r}_{kj}$ is less than $\pi / 2$ radians. Or equivalently the inequality
\begin{equation} 
\mathbf{r}_{ij}\cdot\mathbf{r}_{kj} > 0\;,
\label{CH2Eq:direct_neighbor}\end{equation}
\noindent holds $\forall j$, where $|\mathbf{r}_{ij}|<|\mathbf{r}_{ik}|$. Figs.\ \ref{figCrossFace}(b) and (c) show cases where this inequality \textit{is not} and \textit{is} satisfied respectively.

\begin{figure}
\begin{centering}
\includegraphics[width=8.5 cm]{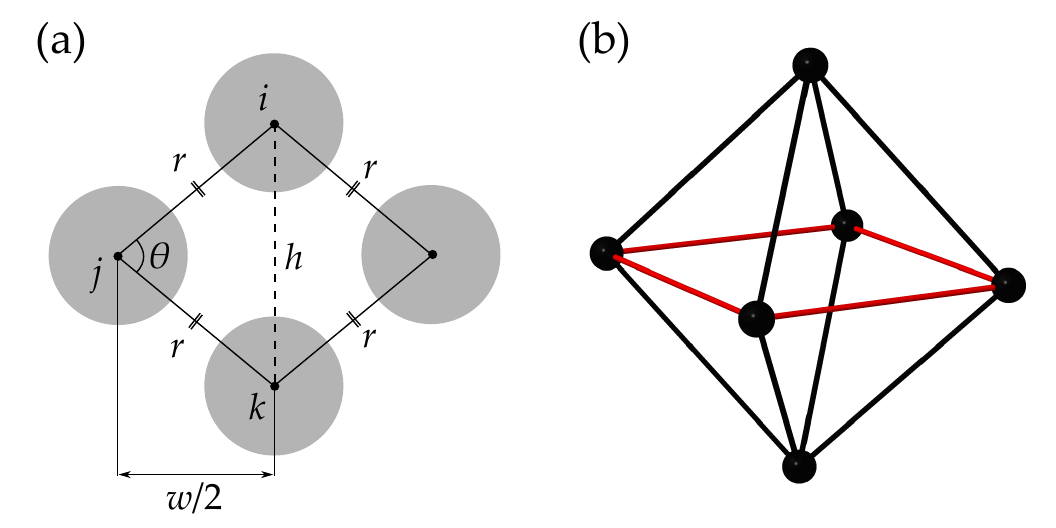} 
\par\end{centering}
\caption{Detection of four-membered rings of particles. (a) Particles are positioned at the vertices of a rhombus. In the case that $h=w$, i.e.\ a square, there is no bond between particles $i$ and $k$. If $h<w$ then a bond forms between $i$ and $k$ across the rhombus if using the standard Voronoi method. For a given rhombus defined by $h/w$ or $\theta$, the existence of the bond between $i$ and $k$ in the modified Voronoi method depends on the value of $f_\mathrm{c}$. (b) The detection of an octahedral cluster with the CNA and TCC methods relies in the integrity of its four-membered rings, e.g.\ the ring highlighted in red. Thermal fluctuations may cause this structure to be detected incorrectly if using the standard Voronoi algorithm to determine the neighbors of each particle.}
\label{figmod_vor} 
\end{figure}

The second modification to the standard Voronoi method improves detection of four-membered rings of particles. Consider particles placed at the vertices of a rhombus, as in Fig.\ \ref{figmod_vor}(a). If the rhombus were a perfect square there would be a total of four bonds detected by the standard Voronoi method, each running along the edges of the square. If the square is distorted by any small amount to form a rhombus, a bond forms between two particles on opposite vertices thus creating two three-membered rings of particles. This bond breaks the detection of any cluster based on the integrity of a four-membered ring, e.g.\ the octahedral cluster shown in Fig.\ \ref{figmod_vor}(b), when any small fluctuation of the particle positions causes a distortion to the ring. 

\begin{figure}
\begin{centering}
\includegraphics[width=8.5 cm]{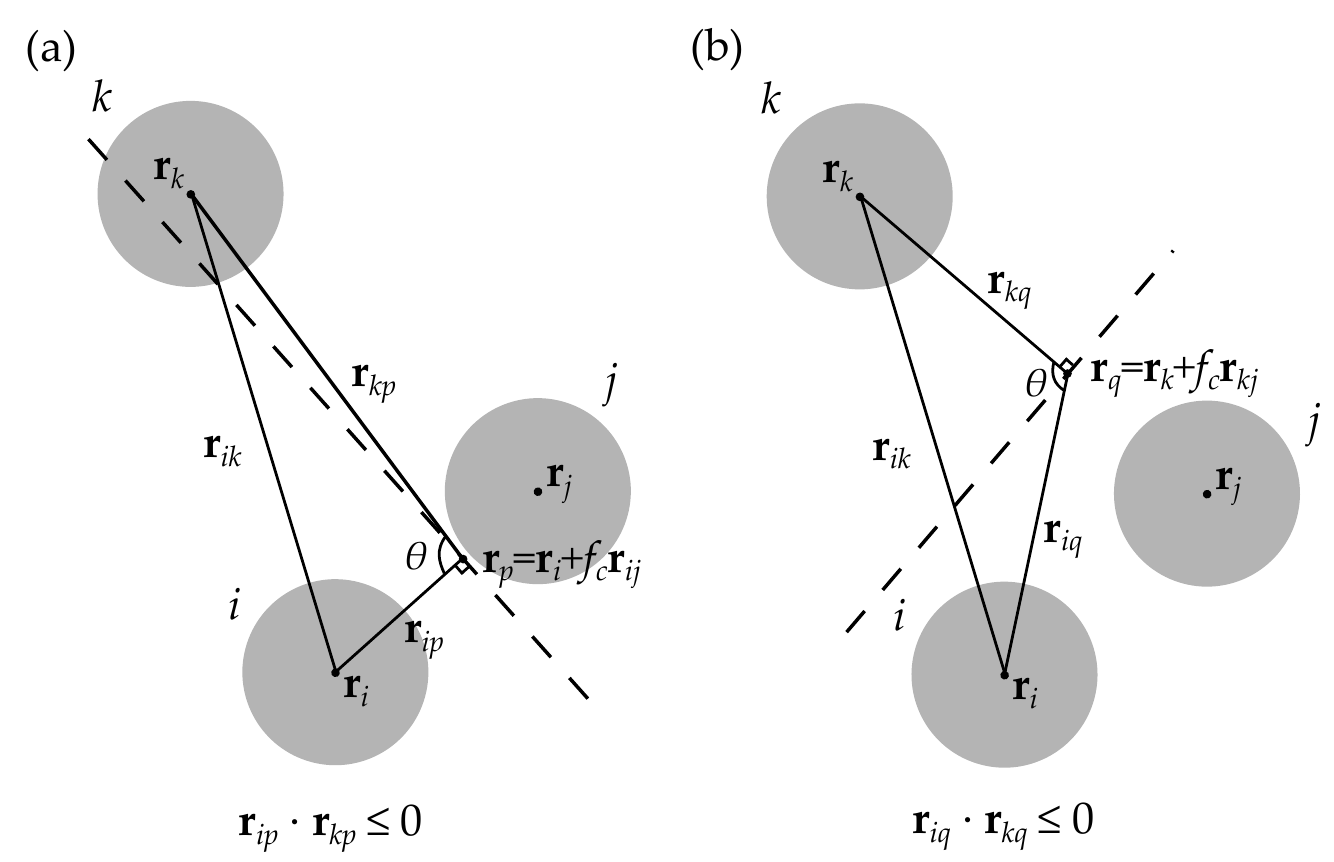} 
\par\end{centering}
\caption[The modified Voronoi method $f_\mathrm{c}$ parameter.]{The modified Voronoi four-membered ring parameter parameter $f_\mathrm{c}$ for the scenario in Fig.\ \ref{figCrossFace}(c). Particles $i$ and $k$ are direct Voronoi neighbors (as per Fig.\ \ref{figCrossFace}(c)), however with the chosen value of $f_\mathrm{c}$ they are not considered as neighbors by the modified Voronoi method. This is because inequality (\ref{CH2Eq:mod_neighbor4}) from the main text is not satisfied. In (a) the particle $k$ lies on the far side of the plane perpendicular to $\mathrm{r}_{ip}$ and containing $\mathrm{r}_{p}$ from particle $i$. Likewise in (b), $i$ lies on the far side of the plane perpendicular to $\mathrm{r}_{kq}$ and containing $\mathrm{r}_{q}$ from $k$. Therefore $\mathbf{r}_{ip}\cdot\mathbf{r}_{kp}+\mathbf{r}_{kq}\cdot\mathbf{r}_{iq}$ must be less than zero.}
\label{mod_vor_method} 
\end{figure}

A dimensionless parameter $f_\mathrm{c}$, known as the four-membered ring parameter \cite{taffs2010}, is introduced that determines the maximum amount of asymmetry that a four-membered ring of particles can display before it is identified as two three-membered rings. Bonds between particles $i$ and $k$ [Fig.\ \ref{figmod_vor}(a)] will be removed if there exists a particle $j$ that is both bonded to $i$ and closer to $i$ than $k$ is, and that shields $k$ from $i$.
We consider the plane perpendicular to $\mathbf{r}_{ij}$ that contains a point $\mathbf{r}_p=\mathbf{r}_i+f_\mathrm{c}(\mathbf{r}_j-\mathbf{r}_i)$. If $f_\mathrm{c}<1$ this corresponds to moving the plane perpendicular to $\mathbf{r}_{ij}$ and containing $\mathbf{r}_j$ towards $\mathbf{r}_i$ [Fig.\ \ref{mod_vor_method}(a)]. Adapting Eq.\ \ref{CH2Eq:direct_neighbor}, the condition for $k$ to be bonded to $i$ is if $\forall j$ where $|\mathbf{r}_{ij}|<|\mathbf{r}_{ik}|$:
\begin{equation} 
\mathbf{r}_{ip}\cdot\mathbf{r}_{kp} > 0\;.
\label{CH2Eq:mod_neighbor}\end{equation}
\noindent Re-expressing the inequality in terms of the position of particle $j$ gives ($f_\mathrm{c} \neq 0$)
\begin{equation} 
f_\mathrm{c}(\mathbf{r}_i\cdot\mathbf{r}_i + \mathbf{r}_j\cdot\mathbf{r}_j - 2\mathbf{r}_i\cdot\mathbf{r}_j) > \mathbf{r}_i\cdot\mathbf{r}_i + \mathbf{r}_j\cdot\mathbf{r}_k -\mathbf{r}_i\cdot\mathbf{r}_j - \mathbf{r}_i\cdot\mathbf{r}_k\;.
\label{CH2Eq:mod_neighbor2}\end{equation}
\noindent Eq.\ \ref{CH2Eq:mod_neighbor2} is not invariant to swapping indices $i$ and $k$. The consequence of this is that $i$ may be bonded to $k$ but not vice versa. It is therefore necessary to consider bonding from the point of view of particle $k$. The plane perpendicular to $\mathbf{r}_{kj}$ containing $\mathbf{r}_j$ is moved towards $k$ such that it contains the point $\mathbf{r}_q=\mathbf{r}_k+f_\mathrm{c}(\mathbf{r}_j-\mathbf{r}_k)$. From the viewpoint of particle $k$, particles $i$ and $k$ are neighbors if $\forall j$ where $|\mathbf{r}_{kj}|<|\mathbf{r}_{ik}|$:
\begin{equation} 
\mathbf{r}_{kq}\cdot\mathbf{r}_{iq} > 0\;.
\label{CH2Eq:mod_neighbor3}\end{equation}

The inequalities in Eqs. \ref{CH2Eq:mod_neighbor} and \ref{CH2Eq:mod_neighbor3} are depicted geometrically in Figs.\ \ref{mod_vor_method}(a) and (b). Both inequalities are not satisfied as the angles $\theta$ are obtuse. 
Adding together Eqs.\ \ref{CH2Eq:mod_neighbor} and \ref{CH2Eq:mod_neighbor3} yields a single, symmetric, criterion for $i$ and $k$ to be bonded. If $\forall j$ with $|\mathbf{r}_{ij}|<|\mathbf{r}_{ik}|$
\begin{equation} 
\mathbf{r}_{ip}\cdot\mathbf{r}_{kp}+\mathbf{r}_{kq}\cdot\mathbf{r}_{iq} > 0\;,
\label{CH2Eq:mod_neighbor4}\end{equation}
\noindent $i$ and $k$ are said to be neighbors in the modified Voronoi method.

Eq.\ \ref{CH2Eq:mod_neighbor4} is invariant to inversion of the indices $i$ and $k$, therefore the resulting modified Voronoi method neighbor network is necessarily symmetric. A symmetric bond network is a required in order to use either the CNA or TCC algorithms to identify structure. Expanding Eq.\ \ref{CH2Eq:mod_neighbor4} in $j$ and setting $f_\mathrm{c}=1$ recovers the definition of all the direct Voronoi neighbors (Eq.\ \ref{CH2Eq:direct_neighbor}). However if $f_c<1$, the modified Voronoi method neighbors (Eq.\ \ref{CH2Eq:mod_neighbor4}) are \textit{only} a subset of the direct Voronoi neighbors.

To demonstrate the effect of $f_\mathrm{c}$ we reconsider the rhombus of particles in Fig.\ \ref{figmod_vor}(a). The parameter $f_\mathrm{c}$ determines the maximum asymmetry of the rhombus ($h/w$) before a bond forms between particles $i$ and $k$ on opposite vertices. Expanding and rearranging Eq.\ \ref{CH2Eq:mod_neighbor4} gives an inequality for a bond to exist between $i$ and $k$ in terms of the particle separations:
\begin{equation} 
f_\mathrm{c} > \frac{r_{ik}^2}{r_{ij}^2+r_{jk}^2}\;.
\label{CH2Eq:four_to_three}\end{equation}
\noindent Using $r_{ij}=r_{jk}$ (side lengths of a rhombus are equal), $r_{ik}=h$  and Pythagoras' theorem gives this condition in terms of the lengths $h$ and $w$:
\begin{equation} 
f_\mathrm{c} > \frac{2}{1+(w/h)^2}\;.
\label{CH2Eq:four_to_three2}\end{equation}
\noindent If $f_\mathrm{c}$ is greater than this value then there is a bond between $i$ and $k$ and the rhombus of particles is identified as two three-membered rings. Conversely if $f_\mathrm{c}$ is less than or equal to this value then $i$ and $k$ are not bonded and the rhombus may be a four-membered ring subject the status of $j$ and $l$. This is found by swapping $w$ and $h$ in Eq.\ \ref{CH2Eq:four_to_three2}.
In terms of the angle $\theta$ subtended by the bonds connecting $j$ to $i$ and $k$, the condition for $i$ and $k$ to be bonded is
\begin{equation} 
f_\mathrm{c} > \frac{2}{1+[\tan(\theta/2)]^{-2}}\;.
\label{CH2Eq:four_to_three3}
\end{equation}
\noindent For angles $\theta$ less than that where Eq. \ref{CH2Eq:four_to_three3} is an equality, $i$ and $k$ are bonded. If $f_\mathrm{c}=1$, the angle $\theta=\pi/2$ and for $f_\mathrm{c}=0.82$ we have $\theta \approx 1.39$ ($\approx 80^{\circ}$). 
The limit for validity of Eq.\ \ref{CH2Eq:four_to_three3} is $\theta\le\pi/3$ ($\theta\le60^{\circ}$), i.e.\ valid choices for $f_\mathrm{c}$ are in the range $(0.5,1]$.

\begin{figure}
\begin{centering}
\includegraphics[width=8.5 cm]{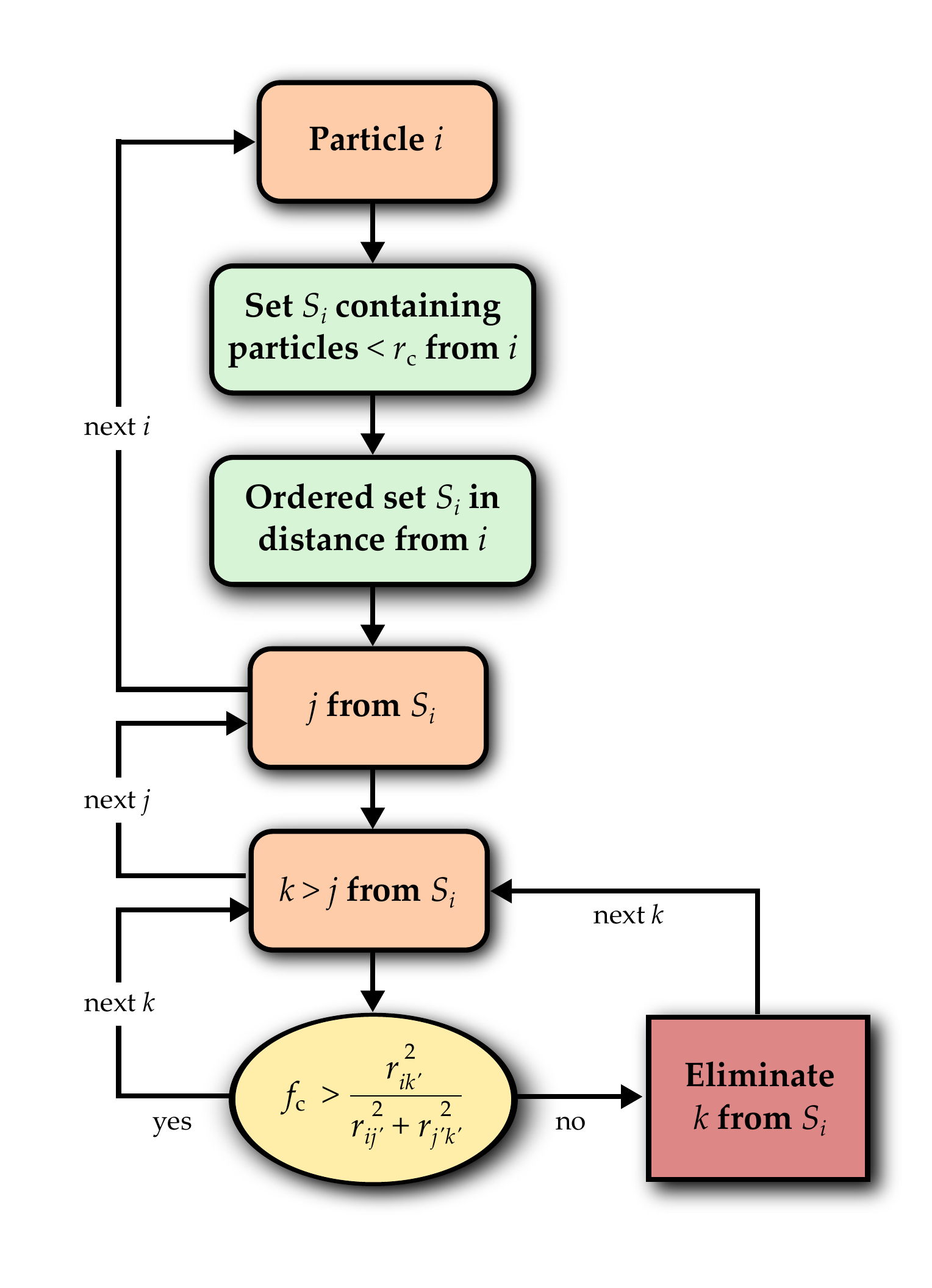} 
\par\end{centering}
\caption{Algorithm for detecting neighbors using the modified Voronoi method. The apostrophes on $j$ and $k$ indicate the particles referenced by the $j^\mathrm{th}$ and $k^\mathrm{th}$ elements of the set $S_i$.}
\label{figmod_vor_algorithm} 
\end{figure}

The algorithm to identify the modified Voronoi neighbor network is therefore as follows (see Fig.\ \ref{figmod_vor_algorithm}):

\begin{enumerate}
  \item Loop over all $N$ particles with index $i$.
  \item Find all particles within $r_\mathrm{c}$ of $\mathbf{r}_i$, where $r_\mathrm{c}$ is to be longer than the longest bond in the network, and add to a set $S_i$. 
  \item Order the particles in $S_i$ by increasing distance from particle $\mathbf{r}_i$.
  \item Loop over all elements $j$ in $S_i$, i.e.\ particles in increasing distance from particle $i$.
  \item For each $j$ loop over all $k>j$ in $S_i$ and eliminate $k^\mathrm{th}$ particle from $S_i$ if inequality \ref{CH2Eq:four_to_three} is not satisfied.
\end{enumerate}

The particles left in the sets $S_i$ on completion are the modified Voronoi method neighbors of particle $i$. 
The value of $r_\mathrm{c}$ can be used to set the maximum bond length in the system. Doing so would be useful if studying gels or cluster fluids, for example, where long bonds are not desired between gaseous particles or particles in separate clusters.

\subsection{Comparison of the neighbor detection methods}
\label{sectionComparisonNeighbor}

\begin{figure*}
\begin{centering}
\includegraphics[width=14 cm]{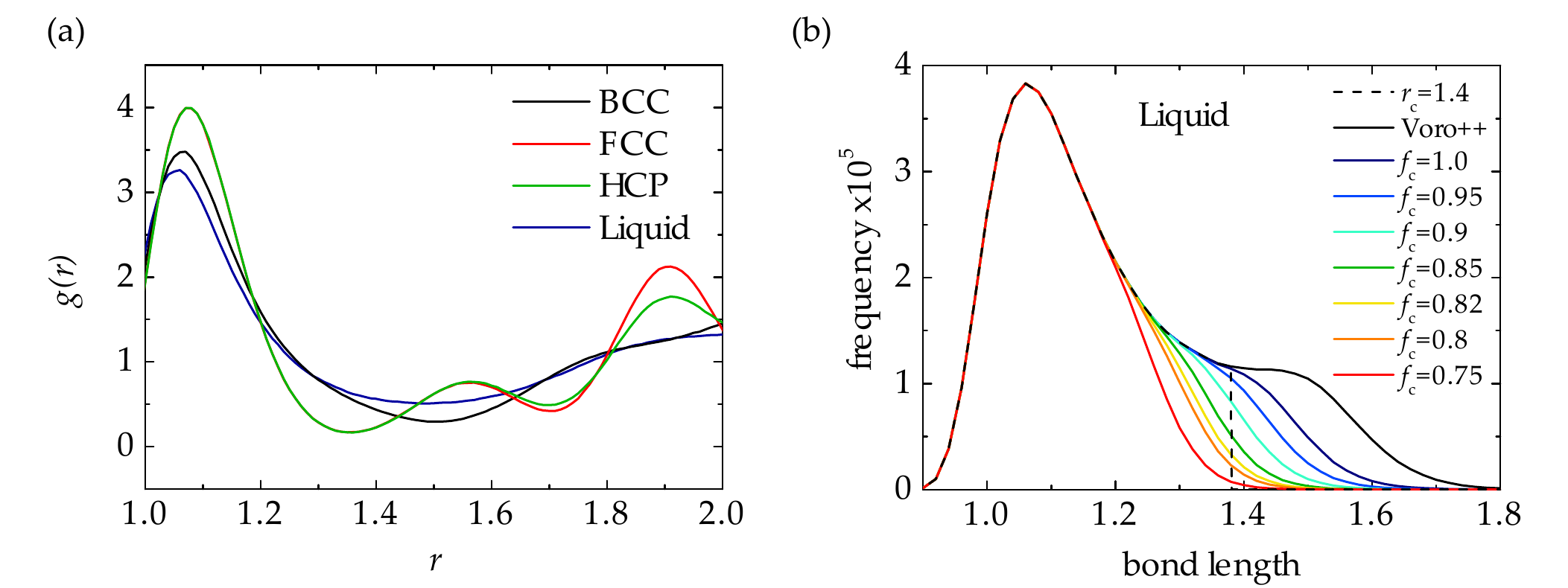} 
\par\end{centering}
\caption{Four phases of a Lennard-Jones system comparing neighbor detection methods. (a) The radial distribution functions for four phases of a Lennard-Jones system at state point $T=0.92$ and $P=5.68$. Note that the position of the first minima of $g(r)$ for the FCC and HCP phases is different from that of the BCC and liquid phases. A single cut-off length cannot determine the neighbors consistently across all the phases. (b) Frequency of bond lengths for different neighbor detection routines. The simple cut-off method is the dotted line ($r_\mathrm{c}=1.4$) and the standard Voronoi method is the black line \cite{rycroft2009}. The colored lines are for the modified Voronoi method. The effect of reducing $f_\mathrm{c}$ is to remove neighbor pairs that have large separations (long bonds) from the standard Voronoi neighbors.}
\label{figLJ_phases} 
\end{figure*}

We compare the neighbor detection methods by considering a Lennard-Jones system at fixed temperature $T=0.92$ and pressure $P=5.68$ as studied by ten Wolde \textit{et al.}\ \cite{tenwolde1996}. Here we use reduced units where Boltzmann's constant $k_\mathrm{B}=1$ and the Lennard-Jones diameter $\sigma=1$.
The simulations are Monte-Carlo in the constant $NPT$ ensemble and the potential is truncated at $r_{\mathrm{tr}}=2.5$. This state point corresponds to roughly $20\%$ undercooling of the liquid phase.
Four phases are considered: a supercooled liquid with $N=864$, a FCC crystal with $N=864$, a HCP crystal with $N=1000$ and a BCC crystal $N=1024$. The supercooled liquid is initialized by random insertion with an overlap separation $r_{\mathrm{ol}}=0.9$, and the crystals by particles on the lattice sites of a crystal filling the simulation box. These samples are used to assess the performance of the TCC in section \ref{sectionResults}, but first we compare the cut-off, standard and modified Voronoi bond detection methods. Before proceeding we note that here the stable phase is the FCC crystal, the others are metastable to varying degrees. However, for our parameters (system size and run time) we saw no change in the state or crystallization to FCC in the metastable states. We choose to simulate one state point in this way so as to provide a fair a test as possible of the TCC's ability to detect different structures.

In Fig.\ \ref{figLJ_phases}(a) the radial distribution function for each phase is plotted about the first minimum. The position of the first minimum for the liquid and BCC phases, $r_{\mathrm{min}}\approx1.5$, is different to that of the FCC and HCP crystal phases, $r_{\mathrm{min}}\approx1.36$. A single cut-off length does not therefore determine the neighbors consistently between the phases with respect to $r_{\mathrm{min}}$. This may cause problems if interfaces exist in the system, or if nucleation and growth is occurring \cite{vanMeel2012}.

In Fig.\ \ref{figLJ_phases}(b) we plot the distribution of bond lengths between neighbors for the under-cooled liquid as identified by the different detection methods. If all particles were neighbors the distribution of bond lengths is an unnormalized plot of $4\pi \rho r^2 g(r)$. All the data coincide up to the first peak indicating that these neighbors are detected by all the different methods for bond detection.
The data for the simple cut-off method drop abruptly to zero at the cut-off length $r_\mathrm{c}=1.4$, whereas the Voronoi methods show a smooth transition to zero. The standard Voronoi method gives the largest number of neighbors in total and includes a number of pairs with separations greater than $r_{\mathrm{min}}=1.5$ for this phase.

The effect of only considering direct Voronoi neighbors as bonded with the modified Voronoi method is clear when comparing the $f_\mathrm{c}=1.0$ data to that of the standard Voronoi method. The difference between the graphs indicates that the indirect Voronoi neighbors have relatively long bonds.
The effect of reducing $f_\mathrm{c}$ is to further reduce the number of neighbors with longer separations. The ideal choice for $f_\mathrm{c}$ will be a function of the structure detection method that is being utilized. Typically cluster detection algorithms, such as the TCC, are sensitive to the inclusion of ``long bonds'' as these may cause misdetection of structures. BOO parameters on the other hand are less sensitive to the inclusion of long bonds providing second nearest neighbors are not included. 
However BOO parameters are themselves sensitive to fluctuations that lead to bonds being broken and reformed \cite{mickel2012}.
A value $f_\mathrm{c}=0.82$ for the TCC method was proposed, based on a study of the hard sphere fluid \cite{williams2007}. We also employ this value in the Lennard-Jones test case in section \ref{sectionTCCexample}. The choice for $f_\mathrm{c}$ is discussed further in the sections below.

\section{The Topological Cluster Classification Algorithm}
\label{sectionTCC}

The topological cluster classification algorithm identifies clusters within a bulk system that are defined by the minimum energy clusters of $m$ isolated particles interacting with a pair-wise potential.
The algorithm works by searching a neighbor network for all three-, four- and five-membered shortest path rings of particles. The rings are categorized by the number of common neighbors to all particles in the ring, forming a set of ``basic clusters''. Larger clusters are then identified as concatenations of the basic clusters and additional particles.

This brief summary of the TCC algorithm allows two important features of the method to be highlighted. (i) Because the minimum energy clusters form the candidate structures that are searched for within the bulk system by the TCC algorithm, in the case that the bulk system is one of the models incorporated in the TCC,
qualitative relationships can be drawn between minimum energy clusters identified and the interactions of the particles in the system. (ii) There is no fixed size for the structural ordering that is considered by the algorithm. 


This section contains a description of the methodology of the TCC algorithm. We proceed to define some pieces of terminology used repeatedly throughout the methodology. 
If two sets of particles are being considered, for example two clusters or two shortest-path rings, we say two particles are \textit{common} if they are members of both sets, and \textit{distinct} or \textit{uncommon} otherwise. \textit{Additional} particles are particles which are not members of the immediate set under consideration.

\subsection{Models}
\label{sectionModels}

\begin{figure}
\begin{centering}
\includegraphics[width=8 cm]{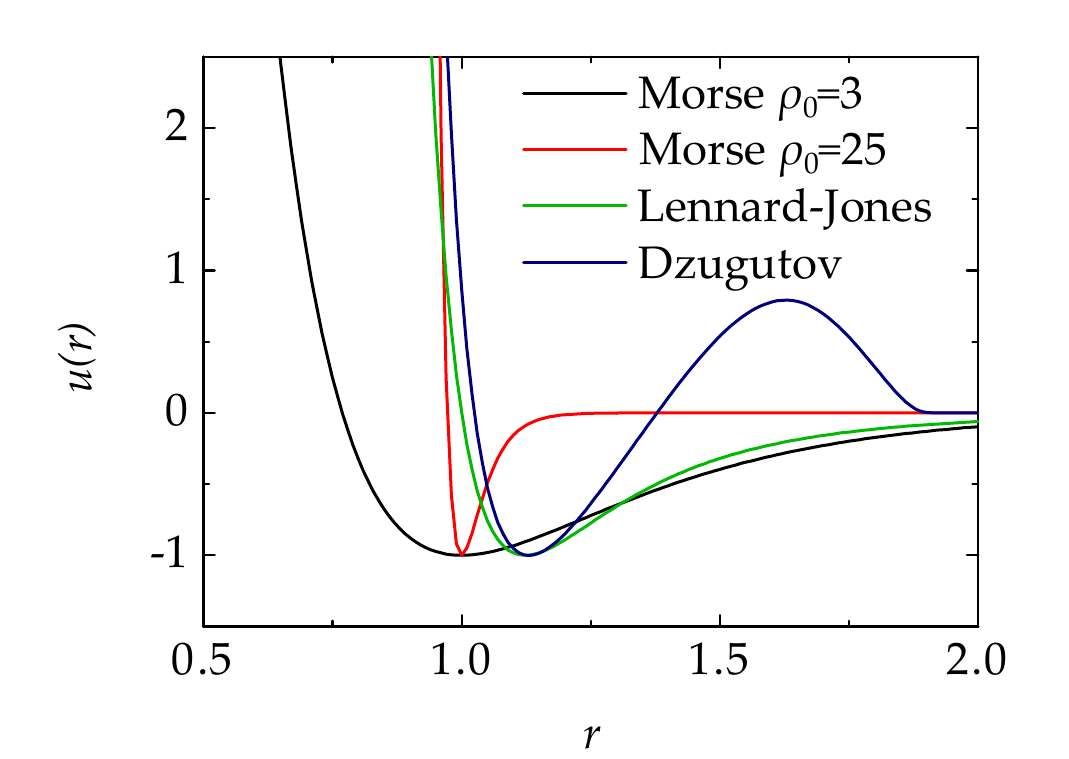} 
\par\end{centering}
\caption{Some potentials considered in the TCC. Black long-ranged Morse ($\rho_0=3.0$), red short-ranged Morse ($\rho_0=25.0$), green Lennard-Jones, and blue is the Dzugutov potential.}
\label{figU} 
\end{figure}

The TCC identifies minimum energy clusters of distinct topology. Thus far, it incorporates a number of models, as described here, but can be adapted to include clusters whose bond topology is not yet included. Models covered include the variable range Morse potential (Fig.\ \ref{figU}),  which reads 

\begin{equation}
u_\mathrm{M}(r)=\varepsilon_\mathrm{M}e^{\rho_{0}(\sigma-r)}(e^{\rho_{0}(\sigma-r)}-2)\;,
\label{eqMorse}
\end{equation}

\noindent where $\varepsilon_\mathrm{M}$ is the depth of the potential well and $\sigma$ is a particle diameter. The range of the potential is set by $\rho_0$ \cite{morse1929,doye1995}. For $\rho_0=6$, the minimum energy clusters for $m\le13$ considered are identical to those of the Lennard-Jones model. We also include the $m=6$ minimum energy cluster of the Dzugutov potential \cite{dzugutov1991,doye2001}.
%
%
%
%
%
As shown in Fig.\ \ref{figU}, the Dzugutov potential is distinguished by a repulsion at around $r=1.5$, introduced to suppress crystallisation.

We also include two binary Lennard-Jones glassformers, the Wahnstr\"{o}m \cite{wahnstrom1991} and Kob-Andersen \cite{kob1995a} models. In the Wahnstr\"{o}m model two equimolar species of Lennard-Jones particles interact with a pair-wise potential,  

\begin{equation}
u_\mathrm{LJ}(r) = 4 \varepsilon_{\alpha \beta}\left[\left(\frac{\sigma_{\alpha\beta}}{r_{ij}}\right)^{12}-\left(\frac{\sigma_{\alpha\beta}}{r_{ij}}\right)^6\right]\;,
\label{eqLJ}
\end{equation}

\noindent where $\alpha$ 
and $\beta$ denote the atom types $A$ and $B$, and $r_{ij}$ is the separation. The (additive) energy, length and mass values are $\varepsilon_{AA}=\epsilon_{AB}=\varepsilon_{BB}$, $\sigma_{BB}/\sigma_{AA}=5/6$, $\sigma_{AB}/\sigma_{AA}=11/12$ and $m_A=2m_B$ respectively. The Kob-Andersen binary mixture is composed of 80\% large (A) and 20\% small (B)
particles possessing the same mass $m$ \cite{kob1995a}. The nonadditive Lennard-Jones interactions between each species, and the cross interaction, are given by
$\sigma_\mathrm{AA}=\sigma$, $\sigma_\mathrm{AB}=0.8\sigma$,
$\sigma_\mathrm{BB}=0.88\sigma$, $\epsilon_\mathrm{AA}=\epsilon$, $\epsilon_\mathrm{AB}=1.5\epsilon$,
and $\epsilon_\mathrm{BB}=0.5\epsilon$.

\subsection{Shortest-path rings}
\label{sectionShortestPathRings}

\begin{figure}
\begin{centering}
\includegraphics[width=8 cm]{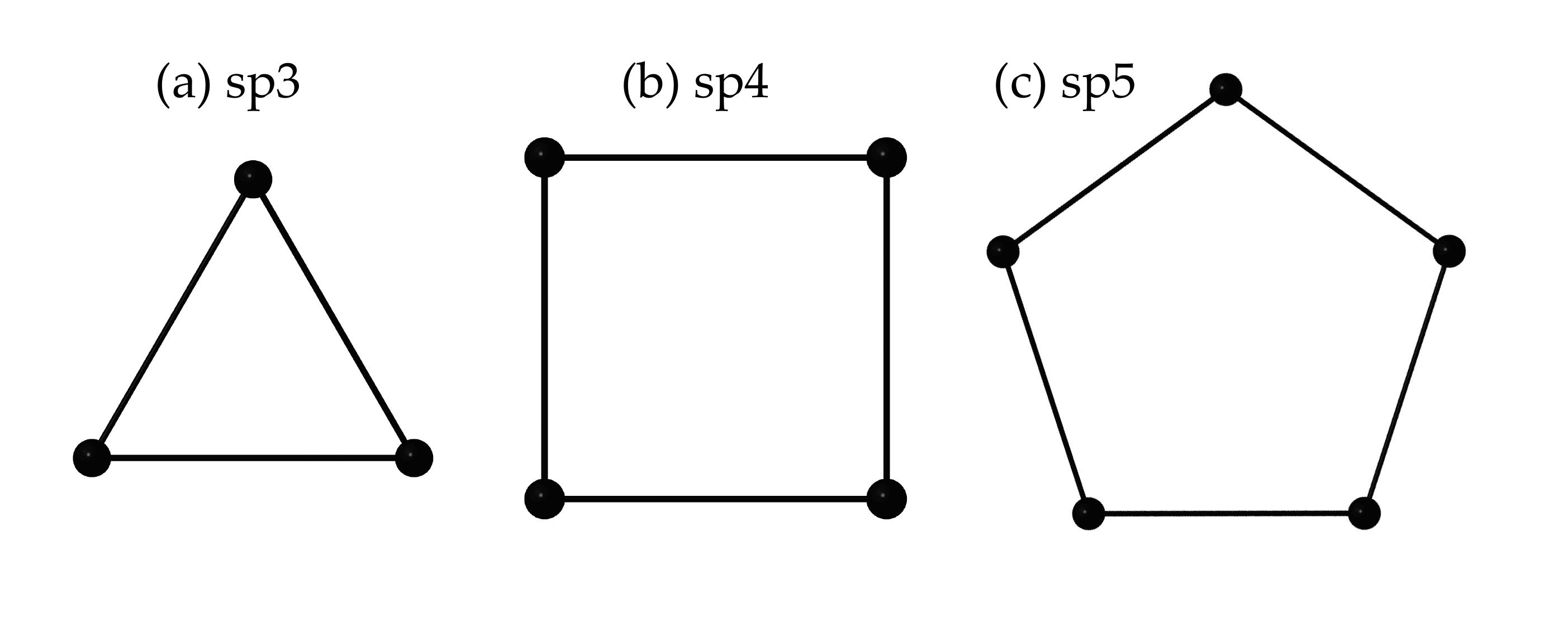} 
\par\end{centering}
\caption{neighbor networks for shortest-path rings of (a) three, (b) four and (c) five particles. Particles are denoted by black spheres and the bonds connecting them by black lines.}
\label{figSP_networks} 
\end{figure}

The starting point of the TCC algorithm is a neighbor network. The algorithm
identifies all the \textit{shortest-path rings} of particles \cite{franzblau1991}. Shortest-path rings are closed loops of bonded particles in the neighbor network where the shortest distance in terms of bonds between any two particles in the ring can be achieved by traversing only bonds between the ring particles. Using more formal notation, if the neighbor network is a graph $G$ consisting of vertices $V$ representing the particles and edges $E$ representing bonds between them, a shortest-path ring is defined as a subgraph $g \subseteq G$ where for all particles $i$ and $j$ within $g$ the equation $d_g(i,j)=d_G(i,j)$ holds ($d_{g/G}(i,k)$ is the minimum number of bonds connecting $i$ and $j$ within the graph). The TCC method considers shortest-path rings of three, four and five particles. The neighbor networks for these shortest-path rings are shown in Fig.\ \ref{figSP_networks}. The three-, four- and five-membered shortest path rings are denoted sp3, sp4 and sp5 respectively.

Shortest-path rings consisting of six or more particles are rarely found in systems with short-range isotropic interactions, unless the particle size disparity is rather higher than the systems we consider (section \ref{sectionModels})  \cite{miracle2009}. Thus so far the TCC is limited to five-membered rings. It would be possible to extend the algorithm for systems with larger shortest-path rings, such as network liquids and glasses, and patchy particles.

\begin{figure}
\begin{centering}
\includegraphics[width=8.5 cm]{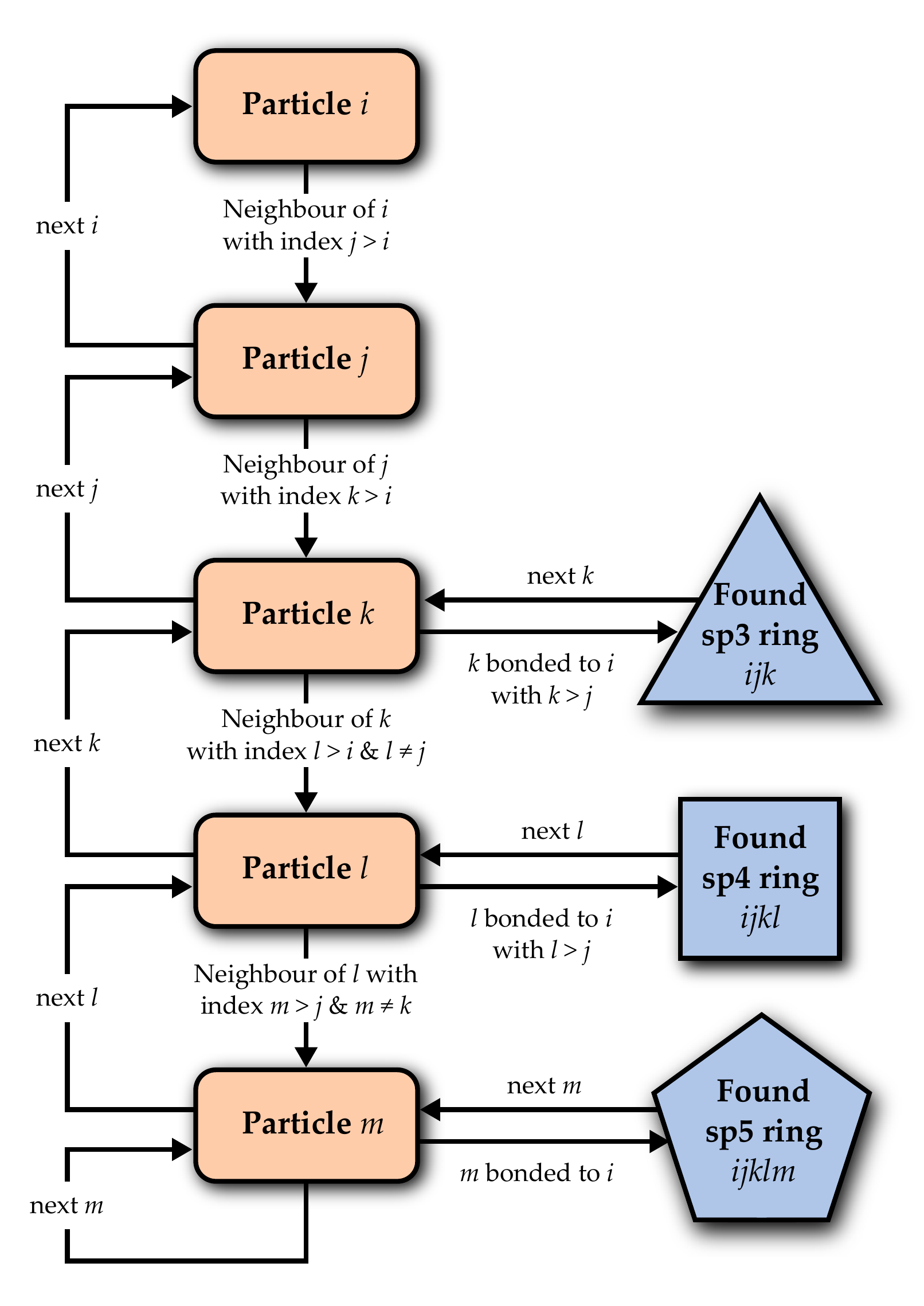} 
\par\end{centering}
\caption{Schematic flow diagram depicting the algorithm used to detect three-, four- and five-membered shortest-path rings.}
\label{figsp_rings_algorithm} 
\end{figure}

The algorithm used to detect the shortest-path rings is similar to that described by Franzblau \cite{franzblau1991} and is depicted in Fig.\ \ref{figsp_rings_algorithm}. Starting from the neighbor network $G$, a loop proceeds over all the particles, index $i$. The three-, four- and five-membered clusters containing $i$ are generated by the backtracking method \cite{nijenhuis1978}. A depth-first search proceeds over the $n_\mathrm{b}(i)$ neighbors of $i$ with index $j>i$. If $j$ has a neighbor that is also bonded to $i$ and with index $k>j$ we have found an sp3 ring. The search then proceeds through the neighbors of $j$ and subsequently back through indices $j$ then $i$. 
If no sp3 ring is identified for index combination $ijk$, the depth-first search proceeds to seek out sp4 rings $ijkl$ where $j<l$, or sp5 rings $ijklm$ where $j<m$. If no sp4 or sp5 ring is found the depth-first search halts and returns to the next neighbor $j$ of $i$. The inequality conditions stated for the particle indices ensure that each ring is detected only once when using this algorithm.
The statistics of shortest-path rings have been used as measures for local structure in their own right, for example in silica glasses \cite{marians1990,franzblau1991} and hard sphere crystals \cite{omalley2001}.

\subsection{The basic clusters}
\label{sectionBasicClusters}

\begin{figure*}
\begin{centering}
\includegraphics[width=14 cm]{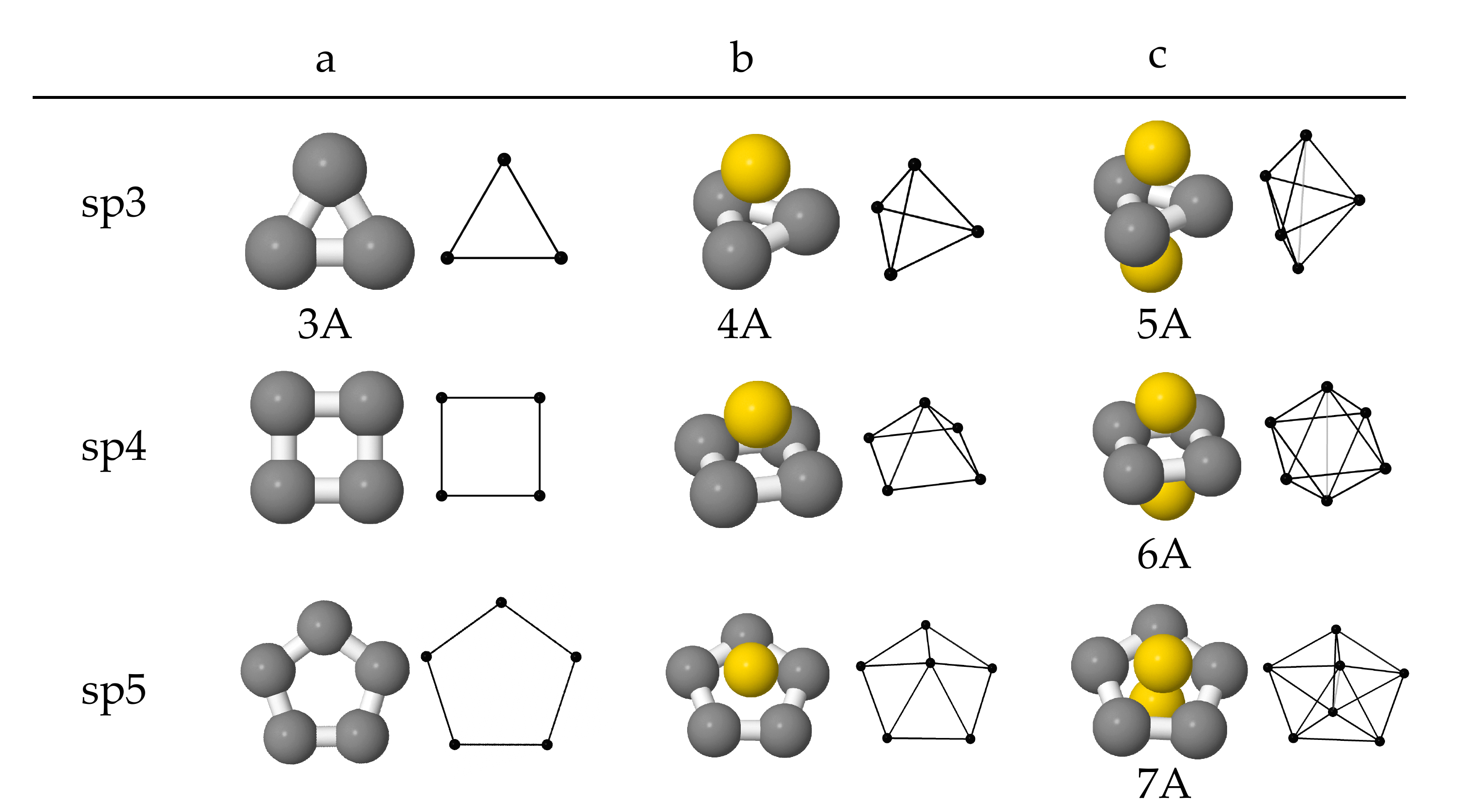} 
\par\end{centering}
\caption{The basic clusters identified from shortest-path rings. If a basic cluster is a minimum energy cluster for $m$ Morse particles its TCC name is shown below the cluster. Two representations are shown for the basic clusters: renderings of the cluster where ring particles are gray, bonds between ring particles are white and spindle particles are yellow, and neighbor networks as in Fig.\ \ref{figSP_networks}.}
\label{figBasicClusts} 
\end{figure*}

The basic clusters are identified upon the shortest-path rings. The shortest-path rings are divided into three further categories, denoted by sp$m$a, sp$m$b or sp$m$c, where $m$ is the number of particles in the ring. For sp$m$a clusters there are no additional particles in the system that are bonded to all $m$ members of the sp$m$ ring. For sp$m$b there is a single additional particle that is a neighbor of all members for the sp$m$ ring. For sp$m$c there are two additional neighbors for all the $m$ ring particles. The additional particles bonded to the ring particles are termed \textit{spindle} particles.

It is assumed that the maximum number of particles that can be bonded to any sp$m$ ring is two, and that if we have an sp$m$c cluster that the two additional particles are located either side of the approximate plane defined by the shortest-path ring particles. These two assumptions are non-trivial and depend on the nature of the interactions, the state point and the method used to derive the neighbor network. The assumptions generally hold true to a good approximation if the interaction potential has a strongly repulsive core at short ranges such that there is a steric minimum distance for approach $r_\mathrm{st}$ between the particles that is in practice not breached. The second requirement is that the maximum length of the bonds is only a small multiple of $r_\mathrm{st}$. 
The validity of these conditions is a function of the interaction potential and the method to detect the neighbors. It is prudent to check that the number of common neighbors to the ring particles is not frequently more than two in order to have confidence that the results obtained from the TCC analysis have physical significance.

For the sp$m$c clusters, no condition is specified as to the bonding between the spindles, i.e.\ they may be bonded or not bonded. In practice the spindles are rarely bonded for sp3c clusters, rarely \textit{not} bonded for sp5c, and bonded in around half of all instances of sp4c clusters.

In total nine basic clusters are defined, as shown in Fig.\ \ref{figBasicClusts}. The basic clusters include the first five minimum energy clusters of the Morse potential consisting of three or more particles. We follow the naming scheme of Doye \emph{et al.}\ \cite{doye1995} and term the minimum energy clusters as 3A trimer (sp3a), 4A tetrahedron (sp3b), 5A triangular bipyramid (sp3c), 6A octahedron (sp4c) and 7A pentagonal bipyramid (sp5c). These are also the minimum energy clusters for the Lennard-Jones model.

\subsection{Compound clusters}
\label{CH2SS:compound_clusters}

Larger clusters are formed of one or more particles bound to one or more basic clusters, or of collections of basic clusters that are bonded to each other. We term these \emph{compound clusters}. Here and in the appendix the detection routines for compound clusters that are identified starting from the basic clusters are described.
For each compound cluster a number of conditions are given for detection. These conditions are expressed in terms of how the constituent particles, either spindle, ring, or additional, are connected by bonds. The conditions are in general satisfied by a number of subtly different bond networks that we term the TCC bond networks for a minimum energy cluster. The TCC bond networks are similar to the bond network of the minimum energy configuration of the cluster. Detecting structure in terms of multiple bond networks allows for some degree of thermally induced distortion of the cluster in the bulk relative to the minimum energy structure.

The aims for a detection routine for a compound cluster are that it is (i) accurate in detecting a given type of order and (ii) simple to implement computationally. If the conditions that a routine imposes for a particular cluster to exist are overly restrictive, thermal fluctuations may mean that detection of the structure in a bulk system is rare. Likewise if the conditions are not sufficiently prescriptive, the order may be detected spuriously or conflicting types of order detected in the same locality.

For the first of the stated aims, the accuracy of detection will be a function of the system under study, the state point and the method used to detect the bonds. The following three general principles are followed relative to the minimum energy bond network in order to maximize the accuracy of the routines:

\begin{enumerate}
	\item The shortest-path rings of particles in basic clusters must lie approximately in a plane in the minimum energy configuration of the cluster.
	\item The bonds that are required for the structure to exist should all be of similar in length in the minimum energy bond network.
	\item The ratio between the shortest bond that if in place would break the detection of the cluster to the longest bond required for the cluster to exist should be maximized (where the bond lengths are taken from the minimum energy configuration).
\end{enumerate}	

\noindent Simplicity of the routines is achieved by utilising the largest sub-clusters where possible from the minimum energy bond network for the cluster. If multiple sub-clusters are going to be required to detect the cluster, the overlap between the sub-clusters should be minimized. These guidelines minimize the number of conditions needed to detect the cluster.

A summary of how the detection routines for the compound TCC clusters are devised is as follows:

\begin{enumerate}
	\item A configuration of particle positions for a cluster is found by identifying a global or local minimum of the potential energy landscape for a given potential. The minimisation is performed using the GMIN method  \cite{wales1997}.
	\item The bond network is found for this configuration using a neighbor detection method. In the case of the Morse, Lennard-Jones and Wahnstr\"{o}m models, this is the modified Voronoi method with $f_\mathrm{c}=0.82$. The value $f_\mathrm{c}=1.0$ is used for the Kob-Andersen model. 
	\item The basic clusters and other smaller compound clusters are identified from within the bond network.
	\item A set of sub-clusters that contain all the particles (perhaps with the inclusion of one or two additional particles) is chosen and the detection routine is based upon these sub-clusters. 
	\item The sub-clusters chosen tend to be the largest available in the bond network, whilst ensuring that the three previously stated principles are obeyed as closely as possible.
	\item Conditions are imposed on the particles shared between sub-clusters, the bonds required between the sub-clusters and additional particles for the structure to exist. The bonds are selected using principles 2 and 3 stated above.
\end{enumerate}

\noindent Detection of clusters is hierarchical: smaller sub-clusters must be detected prior to larger compound clusters.

The naming scheme for the TCC clusters follows from Doye \textit{et al.}\ \cite{doye1995,doye1997} for the minimum energy clusters of the Morse potential. Each name $m$X describes the number of particles in the cluster ($m$) and a character X denotes the potential for which the cluster is a minimum energy configuration. The characters X$=$A,B,C$\dots$ indicate the range $\rho_0$ of the Morse potential for which the cluster is a minimum energy, where A is for the longest ranges, and B, C, $\cdots$ indicate increasingly short range attractions (i.e.\ increasing $\rho_0$). The shortest range Morse potential considered in reference~\cite{doye1997} is $\rho_0=25$.

The 6Z cluster is the minimum energy cluster for six particles of the Dzugutov potential \cite{doye2001} and only a local minimum of the Morse interaction potential for 6 particles. Its detection routine is included as it is a free energy minimum cluster for six colloids with depletion-mediated attractions \cite{arkus2009,malins2009,meng2010}. Minimum energy clusters specific to binary KA and Wahnstr\"{o}m Lennard-Jones potentials are denoted by characters ``K'' and ``W'' respectively.

\subsection{Summary of clusters}
\label{sectioncluster_summary}

\begin{table*}[!ht]															
\begin{centering}															
\begin{tabular}{llccrrrl}															
\hline														
\hline														
Cluster	&	Model	&	$f_\mathrm{c}=0.82$	&	$f_\mathrm{c}=1$	&	$r_\mathrm{l}/r_\mathrm{s}$	&	$r_\mathrm{b}/r_\mathrm{s}$	&	
$r_\mathrm{b}/r_\mathrm{l}$	&	CNA / VFA	\\
\hline																
3A	&	$\rho_0=6$	&	$\bullet$	&	$\bullet$	&	1	&		&		&	~	\\
4A	&	$\rho_0=6$	&	$\bullet$	&	$\bullet$	&	1	&		&		&	~	\\
5A	&	$\rho_0=6$	&	$\bullet$	&	$\bullet$	&	1	&		&		&	CNA-2331	\\
6A	&	$\rho_0=6$	&	$\bullet$	&	$\bullet$	&	1	&	1.41	&	1.41	&	CNA-1441 / 2441	\\
6Z	&	$\rho_0=6$	&	$\bullet$	&	$\bullet$	&	1	&	1.63	&	1.63	&	~	\\
7A	&	$\rho_0=6$	&	$\bullet$	&	$\bullet$	&	1.01	&	1.62	&	1.61	&	CNA-1551	\\
7K	&	KA $m_A=4$	&	(7A)	&	$\bullet$	&	1.26	&	1.78	&	1.41	&	CNA-1551	\\
8A	&	$\rho_0=6$	&	$\bullet$	&	~	&	1.02	&	1.28	&	1.26	&	~	\\
8B	&	$\rho_0=6$	&	$\bullet$	&	$\bullet$	&	1.01	&	1.64	&	1.62	&	~	\\
8K	&	KA $m_A=5$	&	~	&	$\bullet$	&	1.29	&	1.91	&	1.48	&	CNA-1661	\\
9A	&	$\rho_0=6$	&	$\bullet$	&	~	&	1.01	&	1.41	&	1.4	&	~	\\
9B	&	$\rho_0=6$	&	$\bullet$	&	$\bullet$	&	1.01	&	1.62	&	1.59	&	~	\\
9K	&	KA $m_A=6$	&	$\bullet$	&	$\bullet$	&	1.26	&	1.75	&	1.39	&	~	\\
10A	&	$\rho_0=6$	&	$\bullet$	&	~	&	1.01	&	1.41	&	1.4	&	~	\\
10B	&	$\rho_0=6$	&	$\bullet$	&	$\bullet$	&	1.03	&	1.65	&	1.59	&	~	\\
10K	&	KA $m_A=7$	&	$\bullet$	&	$\bullet$	&	1.29	&	1.76	&	1.37	&	~	\\
10W	&	Wahn $m_A=9$	&	$\bullet$	&	$\bullet$	&	1.16	&	1.49	&	1.28	&	VFA-(0,0,9)	\\
11A	&	$\rho_0=3$	&	$\bullet$	&	$\bullet$	&	1.27	&	1.79	&	1.41	&	VFA-(0,2,8)	\\
11B	&	$\rho_0=3.6$	&	$\bullet$	&	~	&	1.18	&	1.3	&	1.11	&	VFA-(0,2,6,2)	\\
11C	&	$\rho_0=6$	&	$\bullet$	&	$\bullet$	&	1.06	&	1.67	&	1.58	&	~	\\
11E	&	$\rho_0=6$	&	$\bullet$	&	$\bullet$	&	1.03	&	1.65	&	1.6	&	~	\\
11F	&	$\rho_0=6$	&	$\bullet$	&	$\bullet$	&	1.01	&	1.42	&	1.41	&	~	\\
11W	&	Wahn $m_A=8$	&	$\bullet$	&	$\bullet$	&	1.24	&	1.78	&	1.44	&	VFA-(0,1,6,3)	\\
12A	&	$\rho_0=2.5$	&	$\bullet$	&	~	&	1.2	&	1.36	&	1.13	&	VFA-(1,0,6,4)	\\
12B	&	$\rho_0=6$	&	$\bullet$	&	$\bullet$	&	1.07	&	1.69	&	1.58	&	~	\\
12D	&	$\rho_0=6$	&	$\bullet$	&	$\bullet$	&	1.04	&	1.66	&	1.6	&	~	\\
12E	&	$\rho_0=6$	&	$\bullet$	&	$\bullet$	&	1.01	&	1.42	&	1.41	&	~	\\
12K	&	KA $m_A=8$	&	$\bullet$	&	$\bullet$	&	1.26	&	1.4	&	1.12	&	~	\\
13A	&	$\rho_0=6$	&	$\bullet$	&	$\bullet$	&	1.05	&	1.7	&	1.62	&	VFA-(0,0,12)	\\
13B	&	$\rho_0=14$	&	$\bullet$	&	$\bullet$	&	1.02	&	1.43	&	1.41	&	VFA-(0,10,2)	\\
13K	&	KA $m_A=7$	&	$\bullet$	&	$\bullet$	&	1.29	&	1.76	&	1.36	&	~	\\
FCC	&	Bulk FCC	&	$\bullet$	&	~	&	1	&	1.41	&	1.41	&	VFA-(0,12)	\\
HCP	&	Bulk HCP	&	$\bullet$	&	~	&	1	&	1.41	&	1.41	&	VFA-(0,12)	\\
9X	&	Bulk BCC	&	$\bullet$	&	$\bullet$	&	1.15	&	1.63	&	1.41	&	~	\\
\hline														
\end{tabular}															
\end{centering}															
\caption{Clusters detected by the TCC algorithm. The second column contains details for the reference potential used to make a configuration of the minimum energy cluster. $m_A$ denotes the number of $A$-species particles in the minimum-energy cluster in the case of the binary systems.
The bullets in the $f_\mathrm{c}$ columns indicate if the TCC algorithm successfully detects the cluster in its minimum configuration. The lengths $r_\mathrm{s}$, and $r_\mathrm{l}$ are the shortest and longest bonds required in the minimum energy bond network in order that the TCC algorithm successfully detects the cluster. The length $r_\mathrm{b}$ is for the shortest bond that could form and result in the cluster no longer being detected. The final column contains suggestions for CNA and VFA clusters with structure similar to the TCC clusters.}
\label{tableTCC}
\end{table*}																

A summary of all the TCC clusters is given in table \ref{tableTCC}. The model used to generate a configuration for a minimum energy cluster is listed. The detection method for each cluster is devised from the bond network which corresponds to the minimum energy cluster in isolation. This bond network necessarily depends on the method used to detect the particles that are neighbors. It is possible that if another neighbor detection method is employed to detect the bond network of the minimum energy cluster, that a different bond network would be obtained. It is therefore not guaranteed that the TCC algorithm would identify a cluster if a neighbor detection method is used that differs from the method used to identify the network for the minimum energy cluster and define the TCC routine for the detection of the cluster.

In practice, if using the simple cut-off or modified Voronoi $f_\mathrm{c}=0.82$, $1$ methods, there are few cases when a cluster would go undetected if using a neighbor network different to that used to define the detection algorithm.
Specifically the Morse clusters 8A, 9A, 10A and 11B fail to satisfy one or more of the conditions for the clusters to exist when additional long bonds are identified using the modified Voronoi method with $f_\mathrm{c}=1$ on the minimum energy configuration. Conversely the KA clusters 7K and 8K are not found using $f_\mathrm{c}=0.82$ as the detection routine relies on the existence of bonds identified when $f_\mathrm{c}=1$.

We consider the stability of the TCC cluster detection to fluctuations in bond lengths by examining the lengths of the bonds in the minimum energy bond networks for each cluster. Specifically we consider the lengths of the shortest $r_\mathrm{s}$ and longest bonds $r_\mathrm{l}$ that are required to exist for the reference minimum energy configuration of a cluster to be correctly detected as that structure. If the ratio $r_\mathrm{l}/r_\mathrm{s}$ is large a bond detection method that includes bonds longer than $r_\mathrm{l}$ is necessary in order to successfully detect the cluster.

As a rule of thumb, the Morse minimum energy clusters have $r_\mathrm{l}/r_\mathrm{s}\approx 1$ meaning that the bonds utilized by the TCC detection routines for these clusters are all of similar length. Exceptions to this rule are the 11A, 11B and 12B clusters, where better detection performance is obtained by using a neighbor detection method that includes bonds with a wider range of lengths, for example the modified Voronoi method with $f_\mathrm{c}\to 1$. The KA and Wahnstr\"{o}m clusters also have relatively large ratios for $r_\mathrm{l}/r_\mathrm{s}$ and a careful choice for the bond detection method is necessary to determine structural order accurately relative to the minimum energy clusters for these systems. We also consider the shortest length $r_\mathrm{b}$ of any additional bond that if included in the bond network would not satisfy one of the conditions for the cluster to exist. The ratio $r_\mathrm{b}/r_\mathrm{s}$ indicates the relative lengthscale of this additional bond.

The ratio $r_\mathrm{b}/r_\mathrm{l}$ contains information on the robustness of the detection routine for each cluster. A large value of $r_\mathrm{b}/r_\mathrm{l}$ indicates that there is a large separation between the longest bond required for the cluster 
to be detected, and the shortest bond that if included results in the cluster going undetected. Clusters with low values of $r_\mathrm{b}/r_\mathrm{l}$, for example 8A, 10W, 11B, 12A, and 12K, are difficult to identify reliably in thermal systems.

Some of the minimum energy clusters with detection routines implemented by the TCC have similar structures to the clusters identified by the CNA and VFA methods. Specifically these are the clusters based around a pair of particles or are a central particle and its complete first coordination shell of neighbors. For these clusters the most similar CNA and VFA clusters are indicated in the final column of table \ref{tableTCC}.

\section{Lennard-Jones FCC and HCP crystals}
\label{sectionTCCexample}

\begin{figure}
\begin{centering}
\includegraphics[width=8 cm]{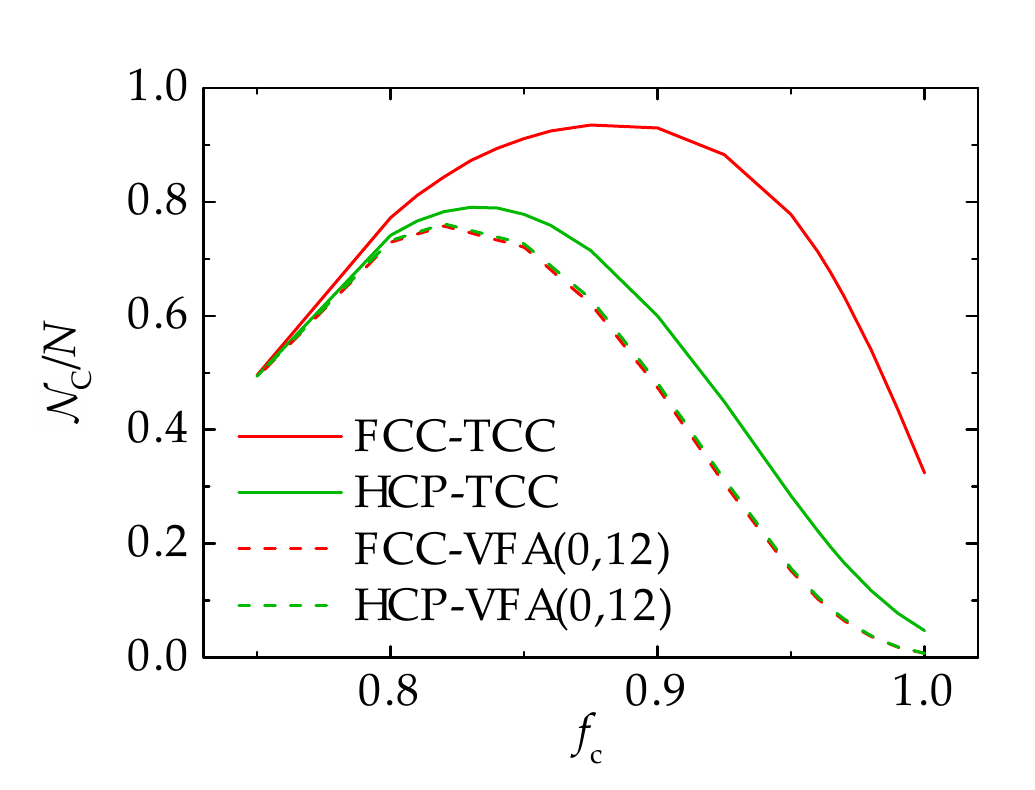} 
\par
\end{centering}
\caption{Comparison of the detection of FCC and HCP crystalline order with the TCC and VFA methods. The number of FCC, HCP and VFA-(0,12) clusters is $\mathcal{N}_\mathrm{C}$.}
\label{figCrystTCCVFA} 
\end{figure}

We demonstrate the TCC algorithm on the simulations of the four Lennard-Jones phases mentioned in section \ref{sectionComparisonNeighbor}. 
Fig.\ \ref{figCrystTCCVFA} shows the ensemble average of the number of FCC and HCP crystal clusters $\mathcal{N}_\mathrm{C}$  versus modified Voronoi parameter $f_\mathrm{c}$ for the FCC and HCP phases. For comparison the number of (0,12) clusters is shown for the two phases as obtained with the VFA method. 
For all values of $f_\mathrm{c}$ the TCC algorithm finds more of the crystal clusters than the VFA method because the TCC detection routines allow for a greater distortion of the order by thermal fluctuations. For both methods the number of clusters tends to zero as $f_\mathrm{c} \to 1$, because of the tendency of the bond detection method to over-estimate the true number of nearest neighbors by including long bonds. The number of clusters identified also tends to zero for low values of $f_\mathrm{c}$ as the bond detection method does not find all of the nearest neighbors.
The optimal value of $f_\mathrm{c}$ for best detection of the crystalline order across both phases and methods is close to the value of $f_\mathrm{c}=0.82$ originally proposed by SRW \cite{williams2007}. 

\begin{figure*}
\begin{centering}
\includegraphics[width=12 cm]{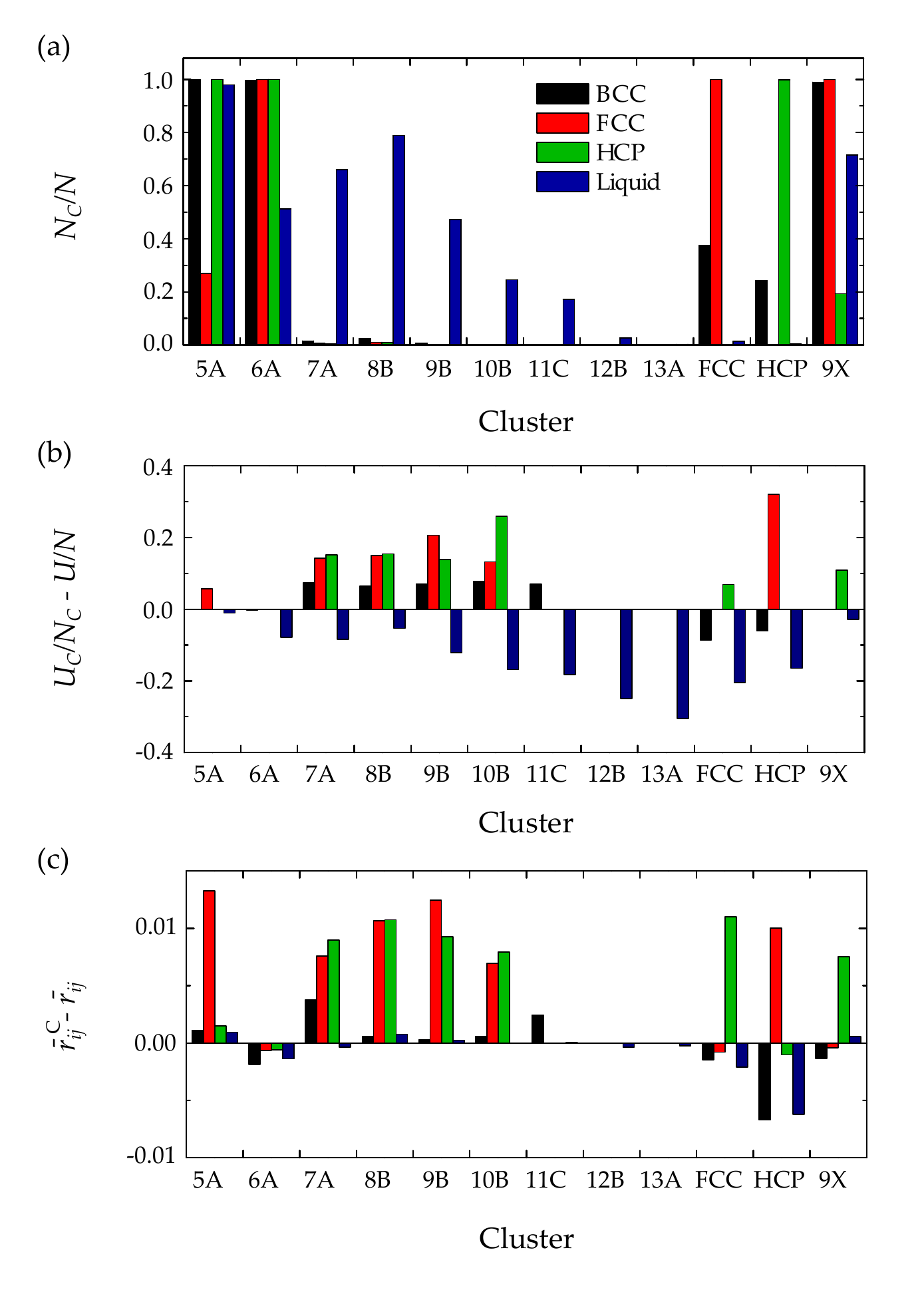} 
\par\end{centering}
\caption[Detection of the LJ minimum energy clusters with the TCC algorithm.]{Detection of the Lennard-Jones minimum energy clusters with the TCC algorithm. The bond detection method is modified Voronoi method with $f_\mathrm{c}=0.82$. (a) Fraction of particles detected within each cluster type. (b) Potential energy difference of particles within each cluster type. (c) Deviation of bond lengths between particles in each cluster type $\bar{r}_{ij}^\mathrm{C}$ from the mean bond length $\bar{r}_{ij}$.}
\label{figLJ_phases_GS} 
\end{figure*}

\section{Four phases of the Lennard-Jones system}
\label{sectionResults}

We proceed to consider the detection of structure in all four phases of the Lennard-Jones system -- BCC, FCC and HCP crystals, and liquid -- for $f_\mathrm{c}=0.82$. Here the state point is the same as that in section \ref{sectionComparisonNeighbor}, thus the stable phase is FCC and the liquid is supercooled.
In Fig.\ \ref{figLJ_phases_GS}(a) the fraction of all particles detected within each cluster type is shown for each phase, where $N_\mathrm{C}$ is the number of particles detected to be within each cluster type (as opposed to the number of clusters $\mathcal{N}_\mathrm{C}$). 

We begin by analysing the detection of crystalline FCC and HCP clusters in their respective bulk phases, as this offers a test for identification of phases with known local structure. All particles are found within FCC and HCP clusters, i.e.\ $N_\mathrm{C}/N\approx 1$, for the respective crystalline phases. We note that $\mathcal{N}_\mathrm{C}\le N_\mathrm{C}$ for FCC and HCP clusters (cf.\ Fig.\ \ref{figCrystTCCVFA}), as the former measure is equivalent to counting the number of particles at the centre of the crystal clusters while the latter also includes particles in the shells of crystal clusters even if they are not themselves at the centre of another crystal cluster.

For the FCC crystal phase, 6A and 9X clusters are detected in large quantities. This result is expected as these clusters form part of the minimum energy structure of the FCC crystal. Similarly 5A and 6A are part of the minimum energy HCP crystal structure. The 5A and other clusters found in trace quantities are not part of the minimum energy FCC crystal structure. These clusters arise due to thermal fluctuations of the particle positions creating transient bonds between particles that do not exist in the minimum energy configuration. The 9X cluster is an example of this phenomenon for the HCP phase.
For the BCC crystal phase 5A, 6A and 9X are structures that are found universally throughout the phase as these are part of the minimum energy crystal structure. Quantities of FCC and HCP are identified along with trace amounts of the other clusters due to thermal fluctuations.

In the liquid phase the 5A structure is most frequently seen and other structures are found in varying quantities. The are trace amounts of FCC and HCP crystalline order, and the 13A icosahedron is the least frequently seen structure. This result is contrary to Frank's original hypothesis \cite{frank1952} and has been shown in earlier studies \cite{stillinger1986,medvedev1987,laViolette1990,taffs2010} and is attributed to 
thermal fluctuations leading to disruption of the bond network, which become more significant for larger clusters. Analysing data following steepest descent quenches to yield inherent structures would therefore be expected to enhance the number of icosahedra identified.
A sizeable fraction of particles within 9X structural environments are found in the liquid phase, confirming the statement in the appendix that this cluster cannot be used to uniquely distinguish crystalline order. Other clusters are seen in various quantities.

In Fig.\ \ref{figLJ_phases_GS}(b) the difference in potential energy between particles identified within each cluster type and the all-particle average is shown. The energy $U_\mathrm{C}$ is the total potential energy of all $N_\mathrm{C}$ particles identified within a certain cluster type. The energy $U_\mathrm{C}/N_\mathrm{C}$ is therefore the mean potential energy of the particles that participate in clusters of type C. This is compared to with the mean potential energy of any particle, $U/N$.

For the FCC and HCP phases the formation of any structural order not found in the minimum energy crystal by thermal fluctuations is associated with an increase in potential energy, relative to the average, for those participating particles. Trace amounts of FCC order are seen in the HCP phase and, although the FCC crystal is the free energy minimum phase for this state point \cite{jackson2002}, the increase in potential energy arises due to the interfacial energy penalty of the boundary introduced between the two crystalline structures.

For the BCC phase the formation of clusters not associated with either BCC, FCC or HCP crystalline order results in an increase in potential energy for the participating particles. When FCC and HCP order is formed the participating particles see a drop in potential energy. This result is an indication of the greater stability of the FCC and HCP phases even at the expense of introducing an interface.

For the liquid phase, the particles that participate in clusters (of any type) have lower potential energy than the mean. The exception is for the 5A structure where almost all the particles participate in that type of cluster ($N_\mathrm{5A}/N\approx 1$), and $U_\mathrm{C}/N_\mathrm{C}$ tends to $U/N$. There is a moderate anti-correlation between $N_\mathrm{C}/N$ and the drop in potential energy for the minimum energy clusters (5A to 13A).

Figure\ \ref{figLJ_phases_GS}(c) displays the effect of participating in clusters on the local packing of particles. The length $\bar{r}_{ij}$ is the mean bond length of the neighbor network, and $\bar{r}_{ij}^{C}$ is the mean length of bonds between particles participating in a particular structure type. The difference $\bar{r}_{ij}^{C}-\bar{r}_{ij}$ indicates the change in bond lengths for particles forming a particular structural order. The FCC and HCP phases show an increase in bond lengths for structures not associated with the pure crystalline order for each phase. This result implies that the formation of any non-crystalline order is associated with local fluctuations of the density. As the mean bond length in uniform systems is proportional to $\rho^{-1/3}$ to a good approximation, the magnitude of any density fluctuation is necessarily small due to the low relative fluctuations in the mean length between the regions of different structure. For the BCC phase the formation of clusters associated with FCC and HCP order results in a local increase in density. There is a local decrease in density for any order found that is not associated with crystallinity.

In the case of the liquid phase only very small changes in the mean bond lengths are seen on formation of the minimum energy clusters 5A to 13A. Again this indicates that the formation of this order is not associated with any local fluctuations in density, but instead caused by different orientational arrangements of the particles being explored. There is an increase in local density on formation of FCC and HCP clusters, however the magnitude of the increase is small compared to the difference in densities of the liquid and crystalline phases at this state point. Related behavior has recently been observed in crystal nucleation in hard spheres, in the formation of a crystal nucleus by structural ordering prior to an increase in density \cite{russo2012}.

\begin{figure*}
\begin{centering}
\includegraphics[width=12 cm]{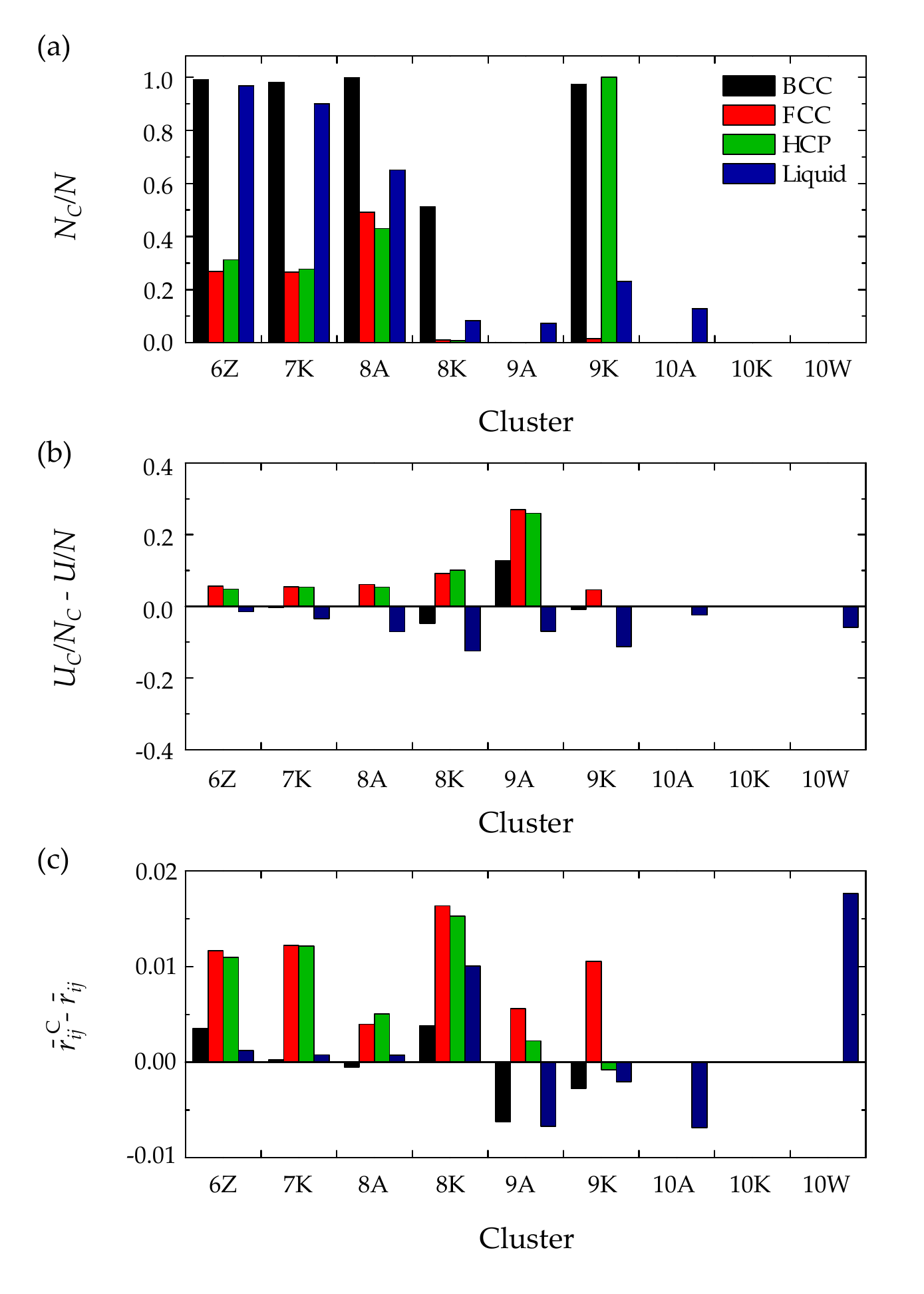} 
\par\end{centering}
\caption[Detection of the other $m=6$ to $10$ particle clusters with the TCC algorithm.]{Detection of the other $m=6$ to $10$ particle clusters with the TCC algorithm. Bond detection method is modified Voronoi with $f_\mathrm{c}=0.82$.}
\label{figLJ_phases_3} 
\end{figure*}

\begin{figure*}
\begin{centering}
\includegraphics[width=12 cm]{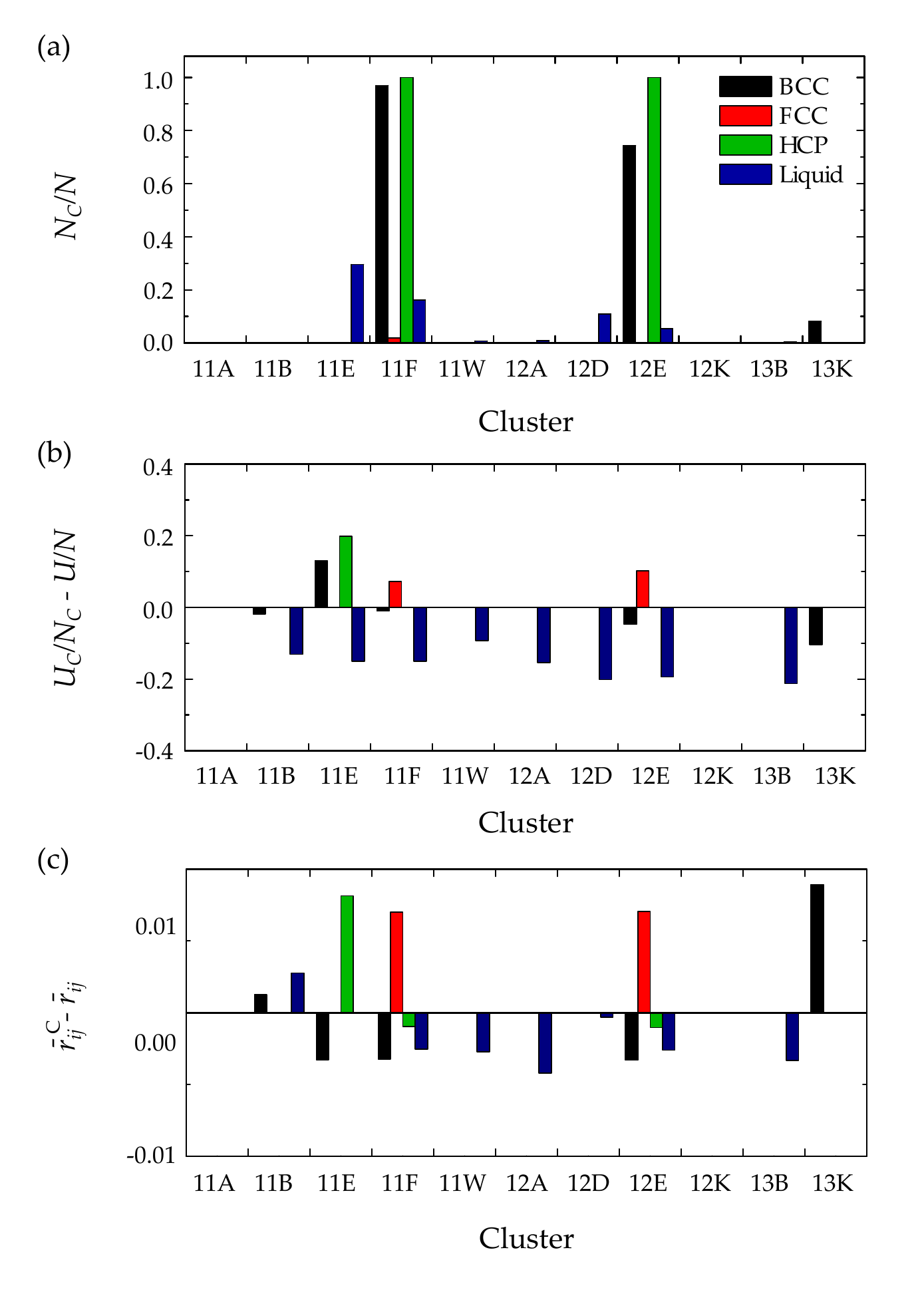} 
\par\end{centering}
\caption{Detection of the other $m=11$ to $13$ particle clusters with the TCC algorithm. Bond detection method is modified Voronoi with $f_\mathrm{c}=0.82$.}
\label{figLJ_phases_4} 
\end{figure*}

For completeness Figs.\ \ref{figLJ_phases_3} and \ref{figLJ_phases_4} contain detection results for all the other structures that are included in the TCC algorithm within the four Lennard-Jones phases. The results are broadly similar to those described above for the minimum energy clusters and the FCC, HCP and 9X clusters. 

\section{Summary}
\label{sectionSummary}

This paper details the topological cluster classification (TCC) algorithm. This method identifies groups of particles whose bond network is similar to that found in minimum energy clusters of a number of monatomic and binary simple liquids. We have described the modification to the Voronoi construction such that 4-membered rings are more robust to thermal fluctuation. These 4-membered rings are required for identification of a number of clusters.
Then, the algorithm was described and its implementation for 33 different clusters was discussed. The TCC was then tested on a bulk one-component Lennard-Jones system in FCC and HCP crystal phases where known local structure was subject to distortion by thermal fluctuations. The test then proceeded to a supercooled liquid phase, where it was identified that particles align into structures with topologies similar to that of a number of minimum energy clusters for the Lennard-Jones and other models. The particles that participate in these structures were found to reduce their potential energy by doing so, and that this occurred not due to density fluctuations but to due to their local changes in arrangement.

\subsection*{Acknowledgements}
We thank Peter Harrowell for introducing C.\ P.\ R.\ to the CNA and Hajime Tanaka for many stimulating conversations. A.\ M.\ is funded by EPSRC grant code EP/E501214/1. C.\ P.\ R.\ thanks the Royal Society for funding. This work was carried out using the computational facilities of the Advanced Computing Research Centre, University of Bristol.


%

\clearpage
\section*{Appendix: Detection Routines for the Clusters}

In the following the symbol $s$ is used to denote the index of a spindle particle in a basic cluster, $r$ to denote a ring particle, and $a$ an additional particle. The subscripts ``$\mathrm{c}$'' and ``$\mathrm{d}$'' on a particle index denote common and distinct particles respectively. Sub-clusters are referenced by the indices $i,j,k,\dots$.

\subsection*{$m=6$ to $m=9$ compound clusters}
\label{CH2SS:10_compound_clusters}

\begin{figure}
\begin{centering}
\includegraphics[width=5.5 cm]{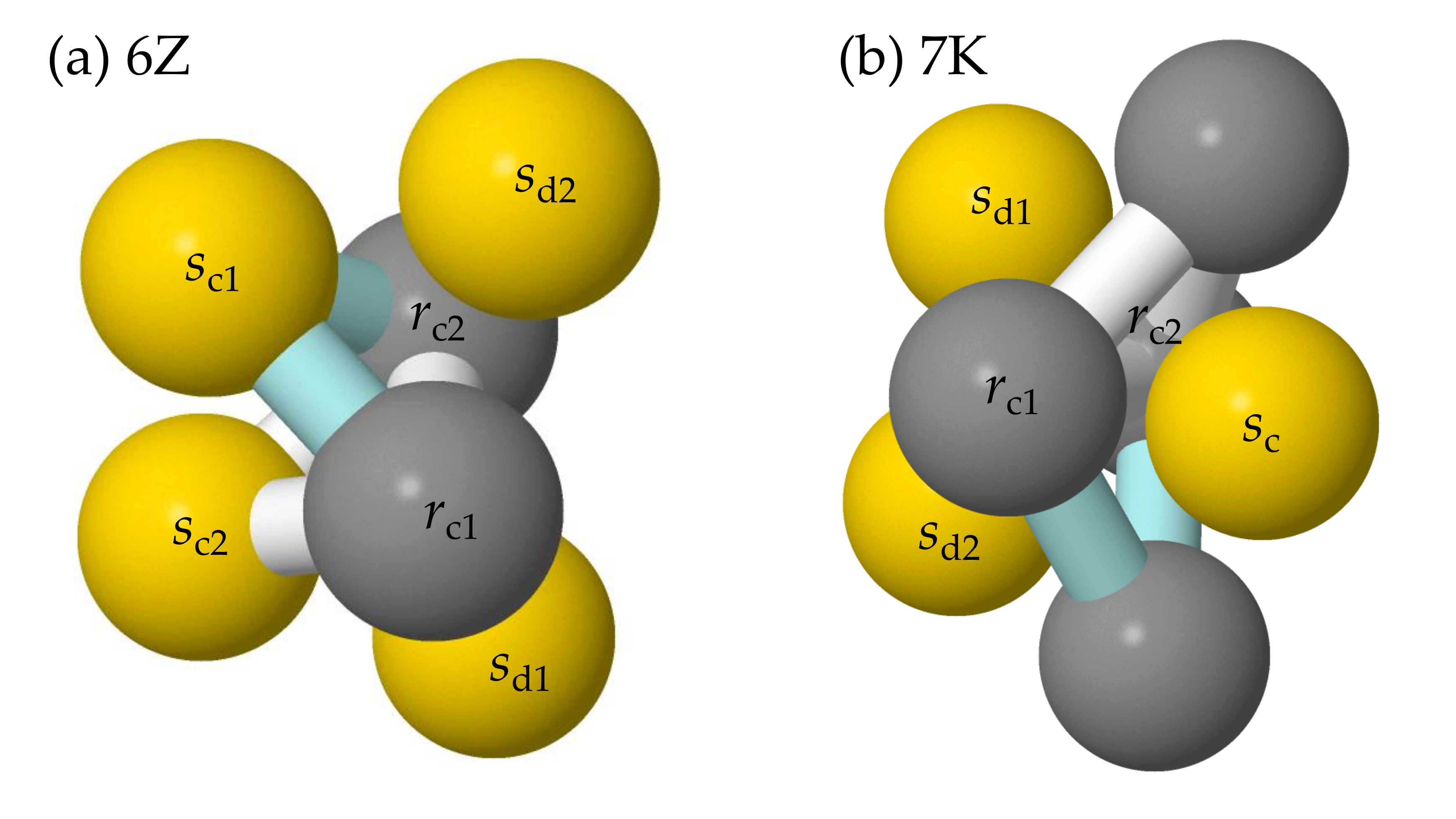} 
\par\end{centering}
\caption[The 6Z and 7A clusters.]{The (a) 6Z and (b) 7A clusters. The two sp3 rings of the 5A clusters are shown in white (5A$_i$) and blue (5A$_j$).}
\label{fig6Z_7K} 
\end{figure}

\textit{6Z. --- } The 6Z cluster is the minimum energy for six particles interacting with the Dzugutov potential \cite{doye2001} [Fig.\ \ref{fig6Z_7K}(a)]. It is also a local minimum of the Morse potential.  See table \ref{tableDetection} for details.

\textit{7K. --- } The 7K cluster is the minimum energy of KA Lennard-Jones for seven particles  [Fig.\ \ref{fig6Z_7K}(b)]. In the minimum energy there are four of the larger $A$-species particles. We give a detection routine based on the bond network of the minimum energy cluster as detected when using the modified Voronoi method with $f_\mathrm{c}=1.0$. The cluster is similar to a 7A cluster, but one of the sp5 ring particles is drawn towards the centre of the ring. The ratio of the longest to the shortest bond in neighbor network of the minimum energy cluster is 1.26. Both these bond types must be in place in the bulk system for the cluster to be detected by the TCC algorithm. See table \ref{tableDetection} for details.

\begin{figure}
\begin{centering}
\includegraphics[width=7 cm]{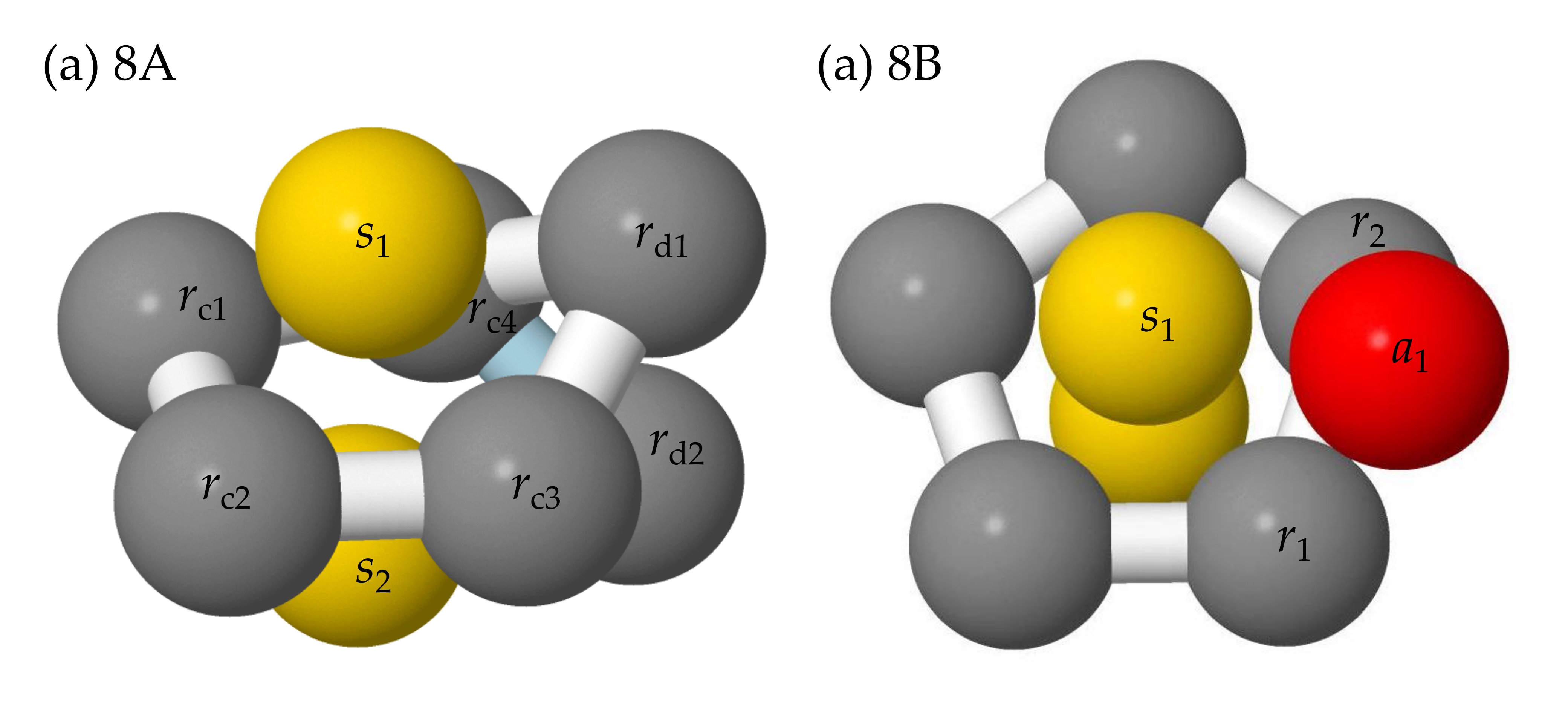} 
\par\end{centering}
\caption[The 8A and 8B clusters.]{The (a) 8A and (b) 8B clusters. For 8A the second sp5 ring is shown in blue. For 8B a red additional particle is bonded to the labeled particles in the 7A cluster.}
\label{fig8A_8B} 
\end{figure}

\textit{8A. --- } The 8A cluster is the minimum energy of the Morse potential for ranges $\rho_0<5.28$  [Fig.\ \ref{fig8A_8B}(a)]. It consists of two sp5b/c clusters and three detection routines are necessary as detailed in Table \ref{tableDetection}.

\textit{8B. --- } The 8B cluster is the minimum energy of the Morse potential for ranges $5.28\le\rho_0<25$ [Fig.\ \ref{fig8A_8B}(b)]. See table \ref{tableDetection} for details.

\begin{figure}
\begin{centering}
\includegraphics[width=7 cm]{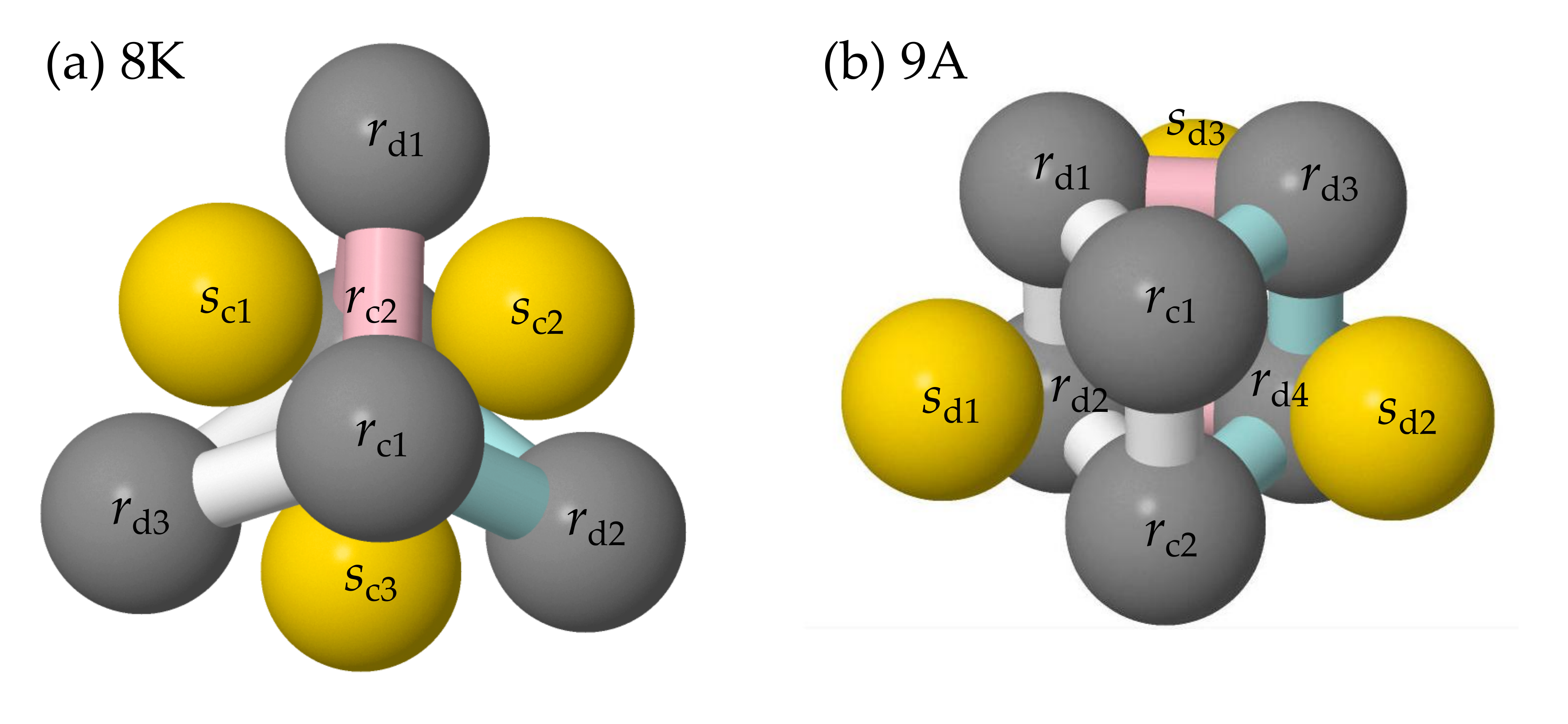} 
\par\end{centering}
\caption[The 8K and 9A clusters.]{The (a) 8K and (b) 9A clusters. Pink bonds indicate the shortest-path ring in the $k$th sub-cluster involved in the detection of 8K and 9A.}
\label{fig8K_9A} 
\end{figure}

\textit{8K. --- } The 8K cluster is the minimum energy of KA Lennard-Jones for eight particles, and contains five of the larger $A$-species particles [Fig.\ \ref{fig8K_9A}(a)]. See table \ref{tableDetection} for details.

\textit{9A. --- } The 9A cluster is the minimum energy for the Morse potential for range parameter $\rho_0<3.42$ [Fig.\ \ref{fig8K_9A}(b)]. See table \ref{tableDetection9} for details.

\begin{figure}
\begin{centering}
\includegraphics[width=7 cm]{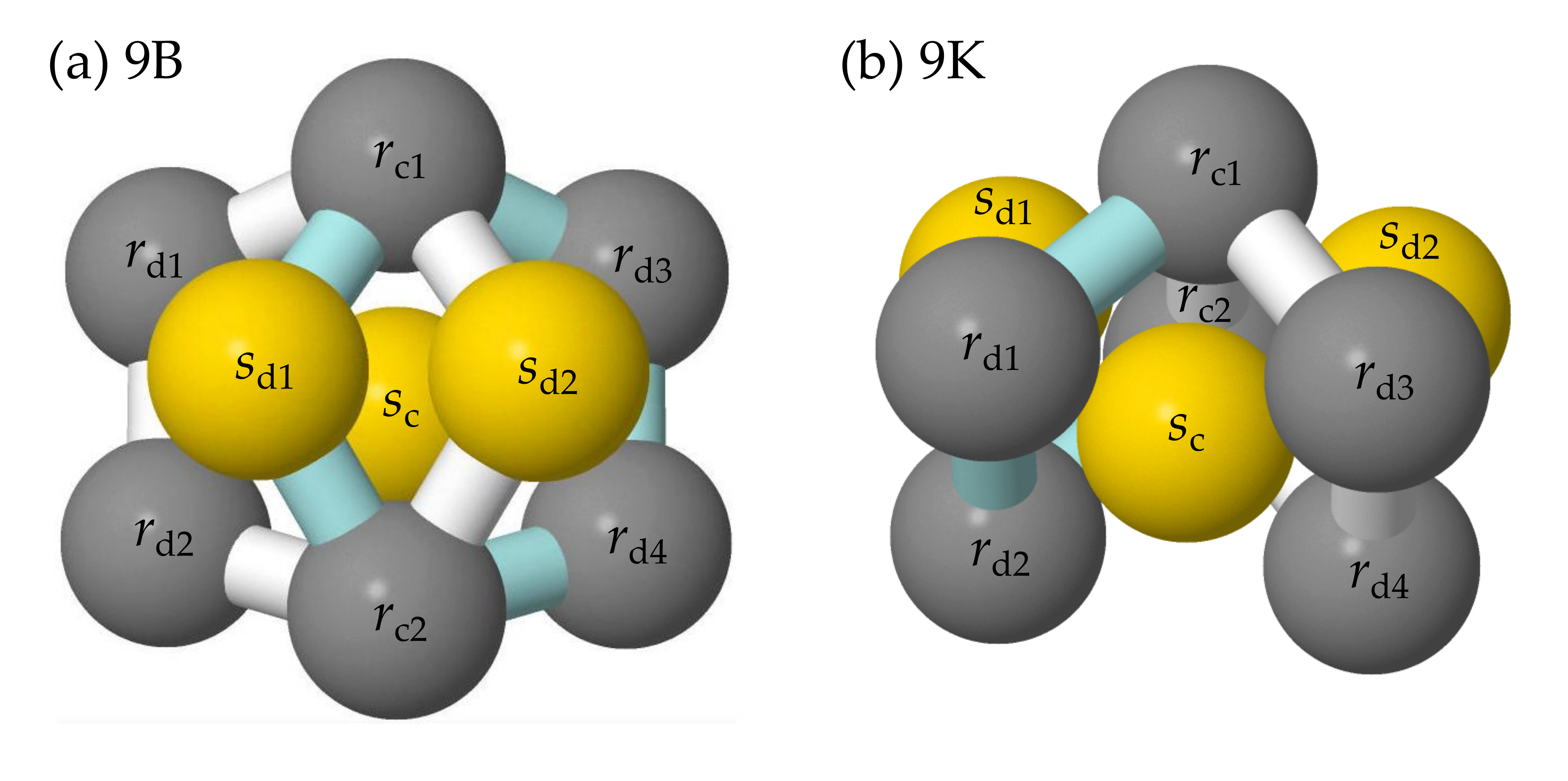} 
\par\end{centering}
\caption[The 9B and 9K clusters.]{The (a) 9B and (b) 9K clusters.}
\label{fig9B_9K} 
\end{figure}

\textit{9B. --- } The 9B cluster is the minimum energy for the Morse potential for ranges $3.42\le\rho_0<25$  [Fig.\ \ref{fig9B_9K}(a)]. See table \ref{tableDetection9} for details.

\textit{9K. --- } The 9K cluster is the minimum energy of KA Lennard-Jones for nine particles, where six of the particles are the larger $A$-species [Fig.\ \ref{fig9B_9K}(b)]. See table \ref{tableDetection9} for details.

\begin{figure}
\begin{centering}
\includegraphics[width=7 cm]{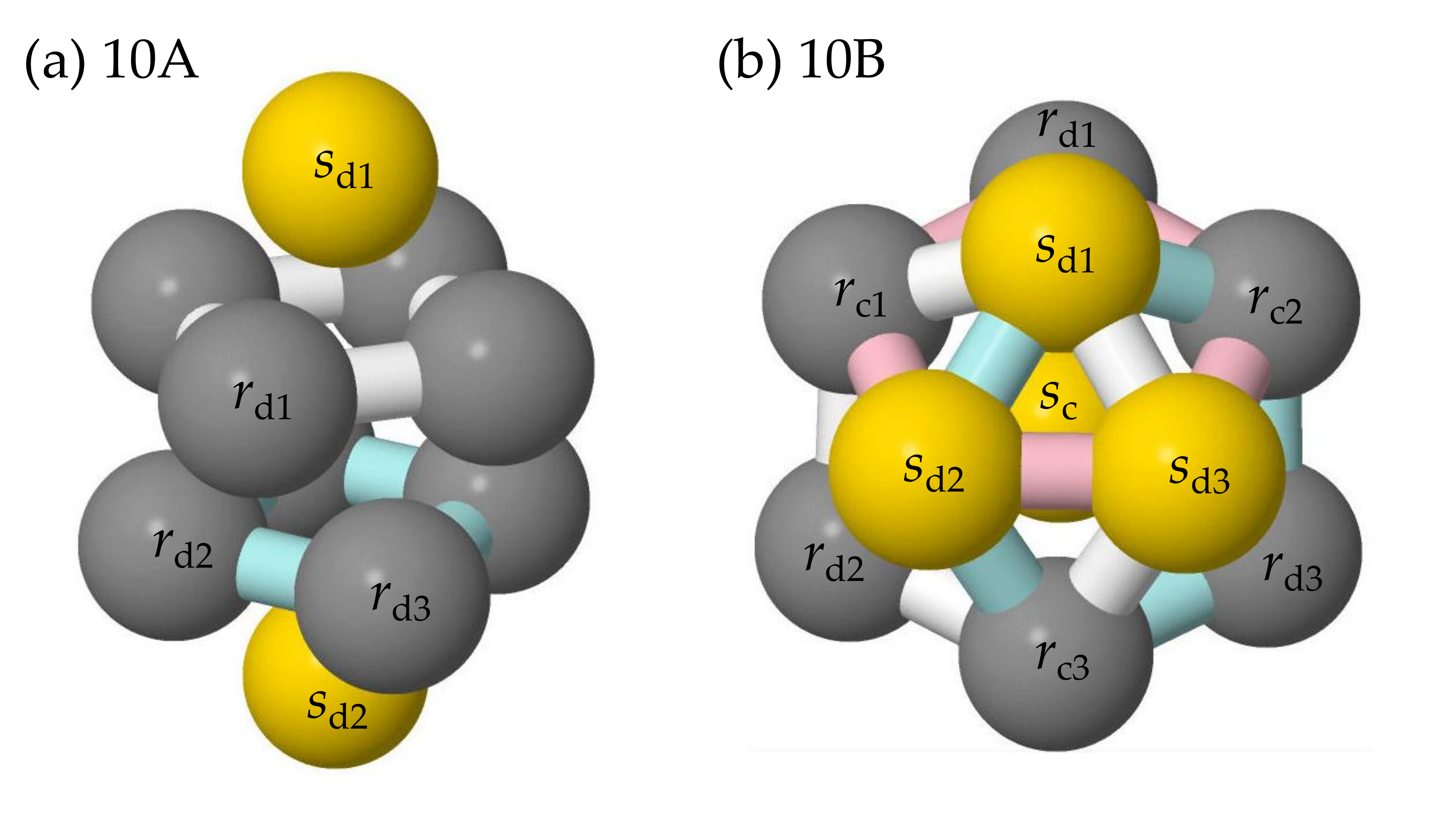} 
\par\end{centering}
\caption[The 10A and 10B clusters.]{The (a) 10A and (b) 10B clusters. For the 10A cluster the three labeled sp4 particles indicate the bonding between the two sp4b clusters.}
\label{fig10A_10B} 
\end{figure}

\textit{10A. --- } The 10A cluster is the minimum energy for the Morse potential for range parameter $\rho_0<2.28$  [Fig.\ \ref{fig10A_10B}(a)]. See table \ref{tableDetection9} for details.

\textit{10B. --- }The 10B cluster is the minimum energy for the Morse potential for ranges $2.28\le\rho_0<25$  [Fig.\ \ref{fig10A_10B}(b)]. The detection routine presented decomposes 10B as a 9B and a 7A cluster. The 10B cluster is the first compound cluster where its detection routine relies on a sub-cluster that is also a compound cluster (9B). It is equally valid to formulate the detection of 10B in terms of three 7A clusters if prior detection of 9B clusters is not of interest. See table \ref{tableDetection9} for details.

\begin{figure}
\begin{centering}
\includegraphics[width=7 cm]{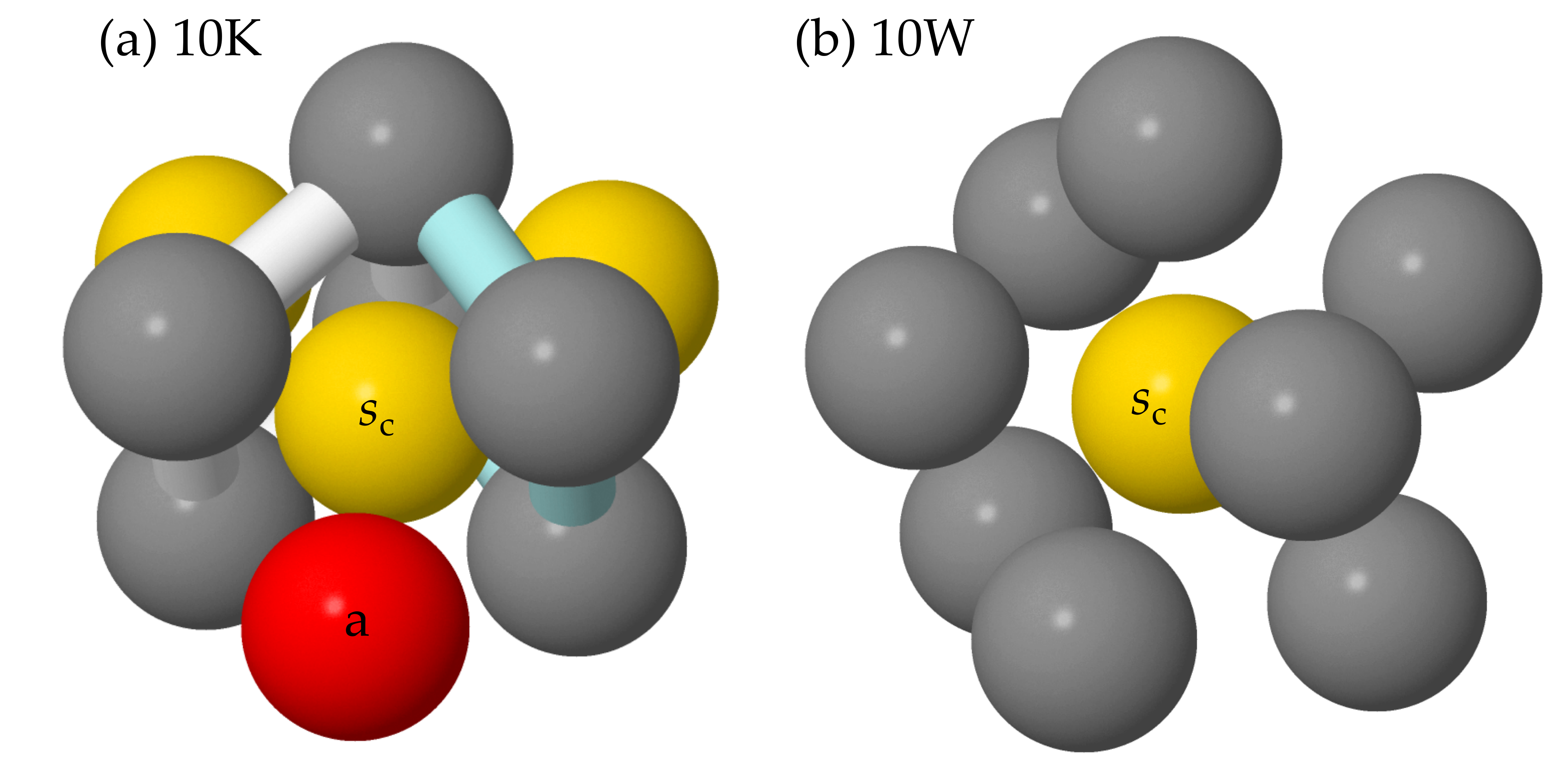} 
\par\end{centering}
\caption{The (a) 10K and (b) 10W clusters. (a) The common spindle particle of 9K and the additional particle are labeled. (b) The common spindle particle of the sp5b clusters constituting 10W is shown.}
\label{fig10K_10W} 
\end{figure}

\textit{10K. --- } The 10K cluster is the minimum energy of KA Lennard-Jones mixture for ten particles, where seven of the particles are the  $A$-species [Fig.\ \ref{fig10K_10W}(a)]. See table \ref{tableDetection9} for details.

\textit{10W. --- } The 10W cluster is only a local minimum of the Wahnstr\"{o}m Lennard-Jones potential for ten particles with nine $A$-species [Fig.\ \ref{fig10K_10W}(b)]. It is included for completeness such that all the energy minimum clusters for each composition of $m_A$ $A$-species for $m$ Wahnstr\"{o}m particles are included in the TCC algorithm. See table \ref{tableDetection9} for details.

\begin{figure}
\begin{centering}
\includegraphics[width = 7cm]{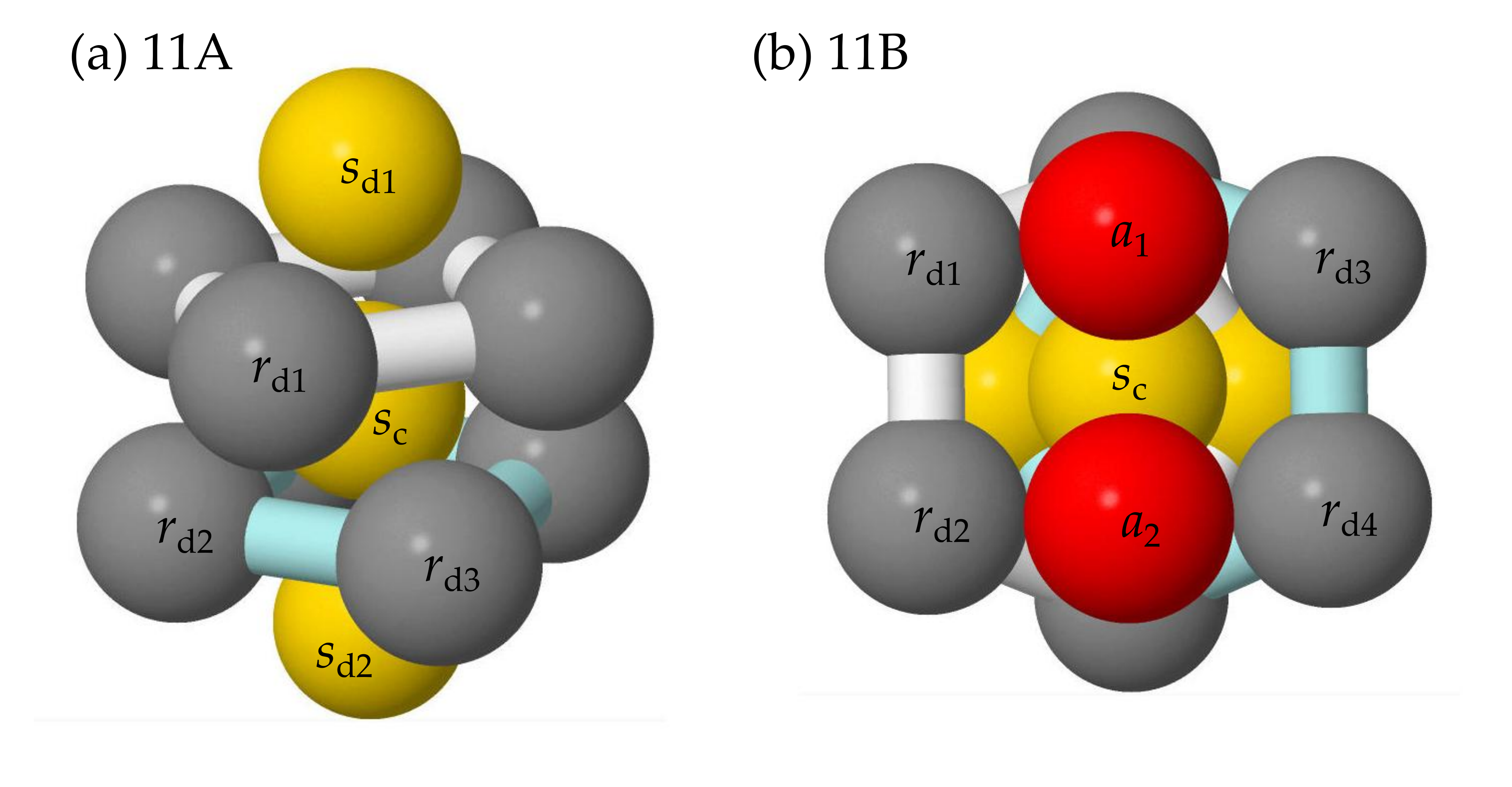} 
\par\end{centering}
\caption[The 11A and 11B clusters.]{The (a) 11A and (b) 11B clusters. (a) The bonding between sp4 ring particles in each 6A is shown by the labeled ring particles. (b) The additional particle $a_1$ is bonded to $r_{\mathrm{d}1}$ and $r_{\mathrm{d}3}$, and $a_2$ is bonded to $r_{\mathrm{d}2}$ and $r_{\mathrm{d}4}$.}
\label{fig11A_11B} 
\end{figure}

\textit{11A. --- } The 11A cluster is the minimum energy for eleven Morse potential for range parameter $\rho_0<3.40$ [Fig.\ \ref{fig11A_11B}(a)]. 
The minimum energy cluster for the Morse potential $\rho_0=3$ is considered to understand the nature of the bonds required to be in place in order that 11A is detected by this routine. The ratio of the lengths of the longest bond to the shortest bond required is 1.27. The longest bonds are those bonds between the sp4 rings of the 6A clusters. The shortest bonds are between the common spindle and the sp4 ring particles. Relatively few 11A clusters are detected when using a bond network that discounts long bonds, e.g.\ when using a short cut-off length or a low value of $f_\mathrm{c}$, as thermal fluctuations cause the bonds between the sp4 ring particles to break and the cluster to go undetected. 

For the KA liquid the 11A cluster has previously been identified as relevant for the glassy behavior \cite{coslovich2007a,speck2012,malins2013fara,royall2014}. The KA minimum energy cluster is topologically equivalent to 11A, i.e.\ it has the same bond network, but the relative lengths are different due to its composition in terms of eight $A$- and three $B$-species. The ratio of the longest to the shortest bonds is 1.23 and this is large enough to cause poor detection of the 11A cluster when using the modified Voronoi method with $f_\mathrm{c}=0.82$. Therefore the value $f_\mathrm{c}=1.0$ is used to determine the neighbors for when studying the KA system \cite{speck2012,malins2013fara}. See table \ref{tableDetection11} for details.

\textit{11B. --- } The 11B cluster is the minimum energy for eleven Morse potential for ranges $3.40\le \rho_0 <3.67$ [Fig.\ \ref{fig11A_11B}(b)]. See table \ref{tableDetection11} for details.

\begin{figure}
\begin{centering}
\includegraphics[width=7 cm]{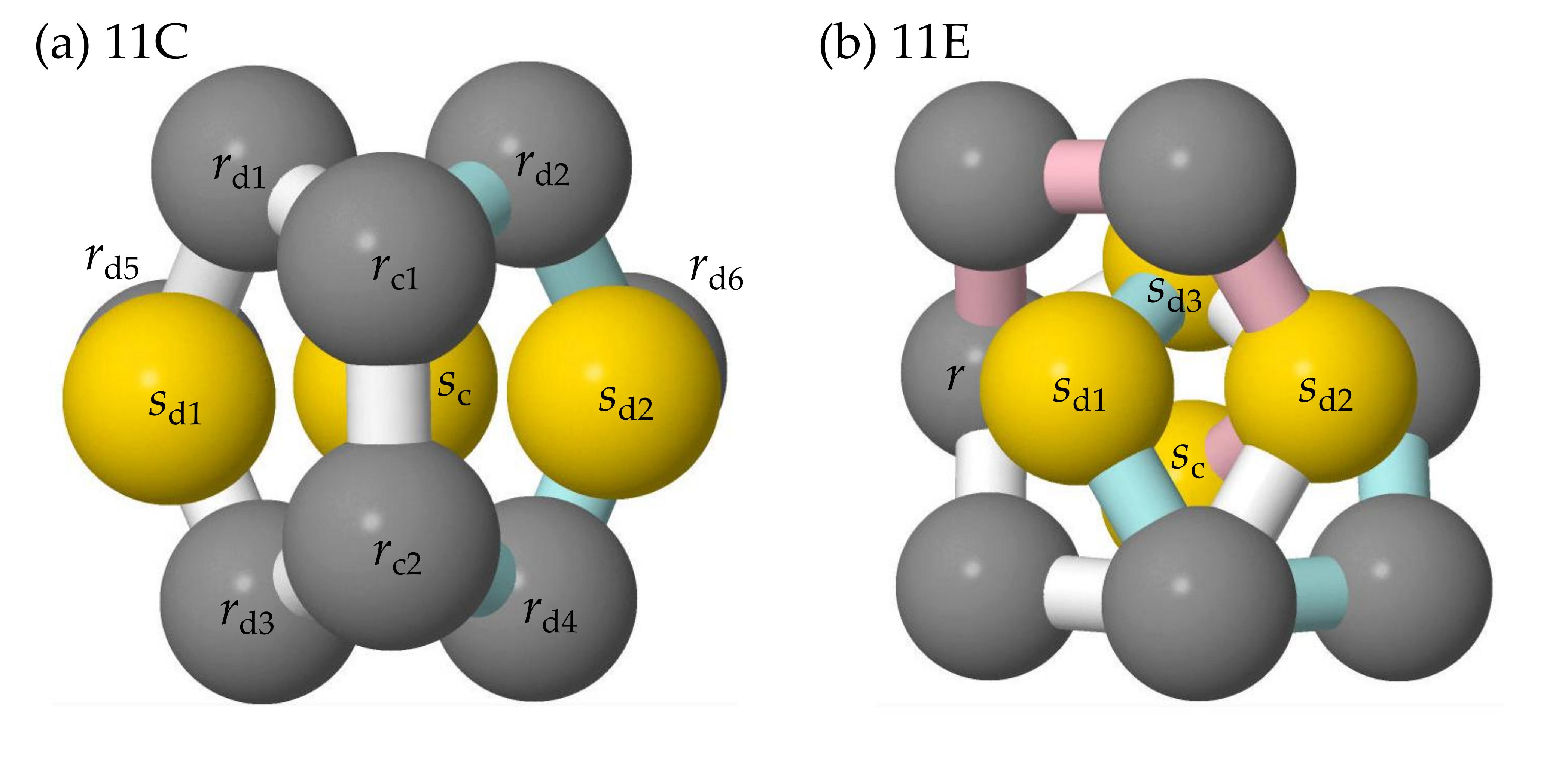} 
\par\end{centering}
\caption[The 11C and 11E clusters.]{The (a) 11C and (b) 11E clusters. (a) Particle $r_{\mathrm{d}1}$ is a neighbor of $r_{\mathrm{d}2}$, and $r_{\mathrm{d}3}$ is a neighbor of $r_{\mathrm{d}4}$. Particles $r_{\mathrm{d}5}$ and $r_{\mathrm{d}6}$ are not neighbors.}
\label{fig11C_11E} 
\end{figure}

\textit{11C and 11D. --- } There are two minimum energy clusters of eleven particles for the Morse potential for ranges $3.67\le \rho_0 <13.57$, namely 11C and 11D [Fig.\ \ref{fig11C_11E}(a)]. Both these clusters have identical minimum energy bond networks, so are not distinguished between by the TCC algorithm. We term both clusters 11C. See table \ref{tableDetection11} for details.

\textit{11E. --- }The 11E cluster is the minimum energy for eleven Morse particles with the range of the potential in $13.57\le \rho_0 <20.60$ [Fig.\ \ref{fig11C_11E}(b)]. See table \ref{tableDetection11} for details.

\begin{figure}
\begin{centering}
\includegraphics[width=7 cm]{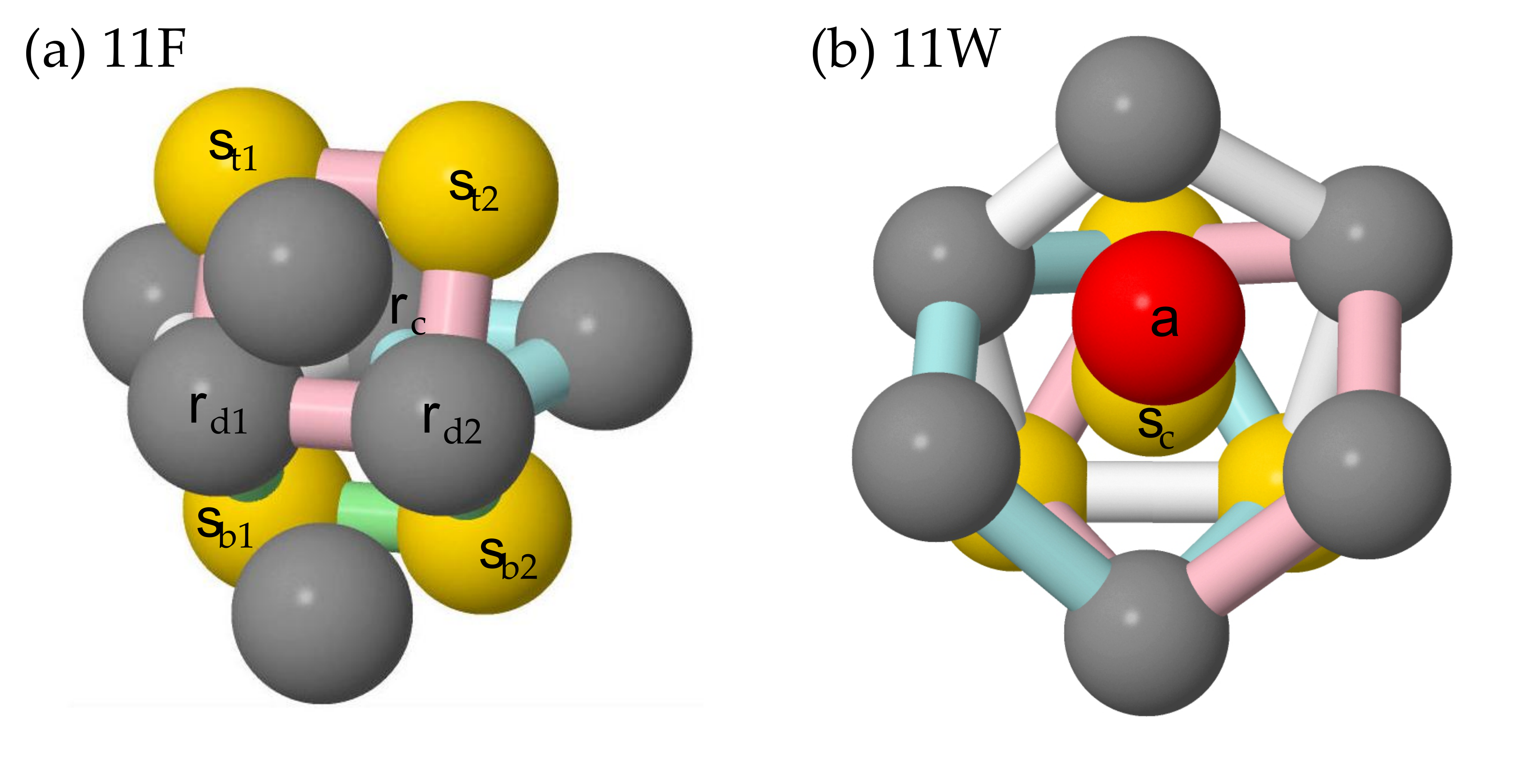} 
\par\end{centering}
\caption[The 11F and 11W clusters.]{The (a) 11F and (b) 11W clusters. (a) The white and blue sp3 rings are indicate the 5A clusters, and the pink and green sp4 rings indicate the 6A clusters. Only the spindles of the 5A clusters are highlighted in yellow. (b) The additional neighbor $a$ of the common spindle $s_{\mathrm{c}}$ in the 10B cluster is highlighted.}
\label{fig11F_11W} 
\end{figure}

\textit{11F. --- }The 11F cluster is the minimum energy for eleven Morse particles with the range of the potential in $20.60\le\rho_0<25$  [Fig.\ \ref{fig11F_11W}(a)]. See table \ref{tableDetection11F} for details.

\textit{11W. --- } The 11W cluster is the minimum energy of Wahnstr\"{o}m Lennard-Jones mixture for eleven particles  [Fig.\ \ref{fig11F_11W}(b)]. Nine of the particles are the larger $A$-species. See table \ref{tableDetection11F} for details.

\begin{figure}
\begin{centering}
\includegraphics[width=7 cm]{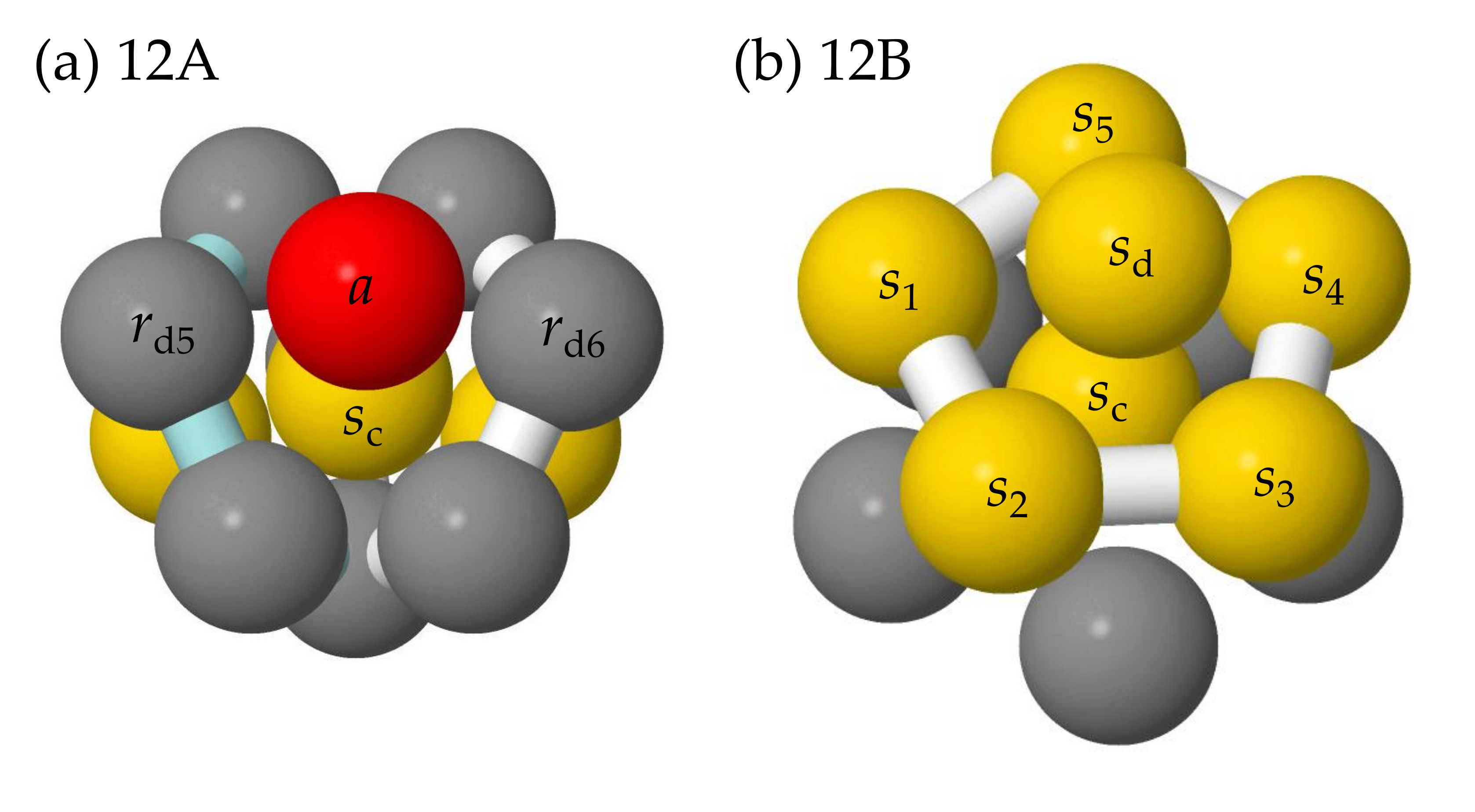} 
\par\end{centering}
\caption[The 12A and 12B clusters.]{The (a) 12A and (b) 12B clusters. (a) The additional particle is bonded only $s_\mathrm{c}$, $r_{\mathrm{d}5}$ and $r_{\mathrm{d}6}$. (b) Five 7A clusters have spindles $s_\mathrm{c}$ and $s_1$ to $s_5$, the latter being bonded to $s_\mathrm{d}$ of the central 7A cluster.}
\label{fig12A_12B} 
\end{figure}

\textit{12A. --- }The 12A cluster is the minimum energy for twelve Morse particles with the range of the potential in $\rho_0 < 2.63$ [Fig.\ \ref{fig12A_12B}(a)]. See table \ref{tableDetection11F} for details.

\textit{12B/C. --- } There are two minimum energy clusters of twelve particles for the Morse potential for ranges $2.63\le \rho_0 <12.15$, namely 12B and 12C [Fig.\ \ref{fig12A_12B}(b)]. Both these clusters have identical minimum energy bond networks and so cannot be distinguished between by the TCC algorithm. We term both clusters 12B. See table \ref{tableDetection11F} for details.

\begin{figure}
\begin{centering}
\includegraphics[width=7 cm]{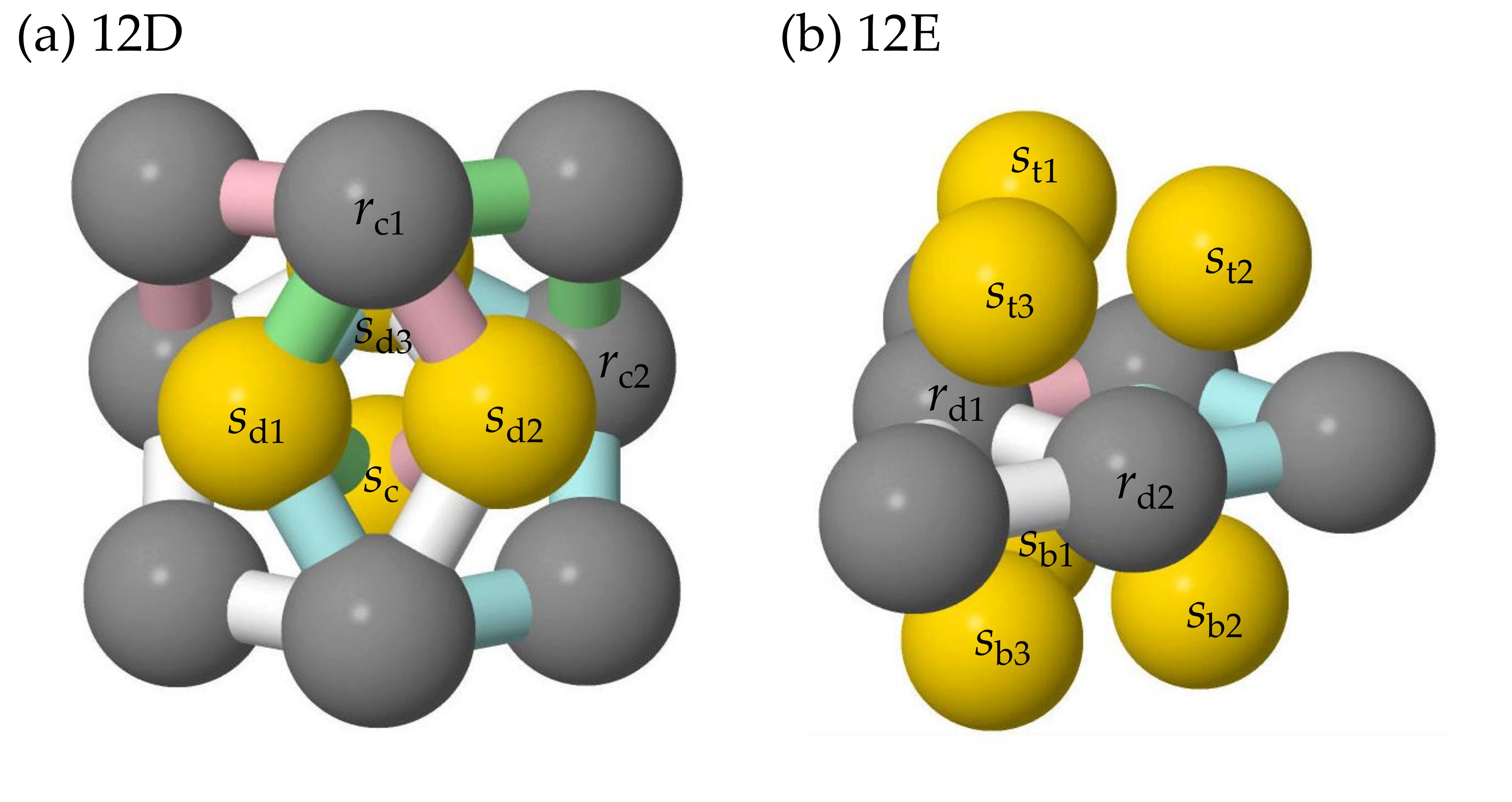} 
\par\end{centering}
\caption{The (a) 12D and (b) 12E clusters. (a) The 7A cluster added to the 11E cluster has its sp5 ring highlighted with green bonds. (b) The additional 5A cluster to the 11F cluster has its sp3 ring highlighted with white bonds and its spindles are $s_{\mathrm{t}3}$ and $s_{\mathrm{b}3}$.}
\label{fig12D_12E} 
\end{figure}

\textit{12D. --- } The 12D cluster is the minimum energy for twelve Morse particles with the range of the potential in $12.15 \le \rho_0 < 17.08$  [Fig.\ \ref{fig12D_12E}(a)]. See table \ref{tableDetection11F} for details.

\textit{12E. --- } The 12E cluster is the final Morse cluster for twelve particles [Fig.\ \ref{fig12D_12E}(b)]. It is the minimum energy for ranges $17.08\le\rho_0<25$. See table \ref{tableDetection11F} for details.

\begin{figure}
\begin{centering}
\includegraphics[width=7 cm]{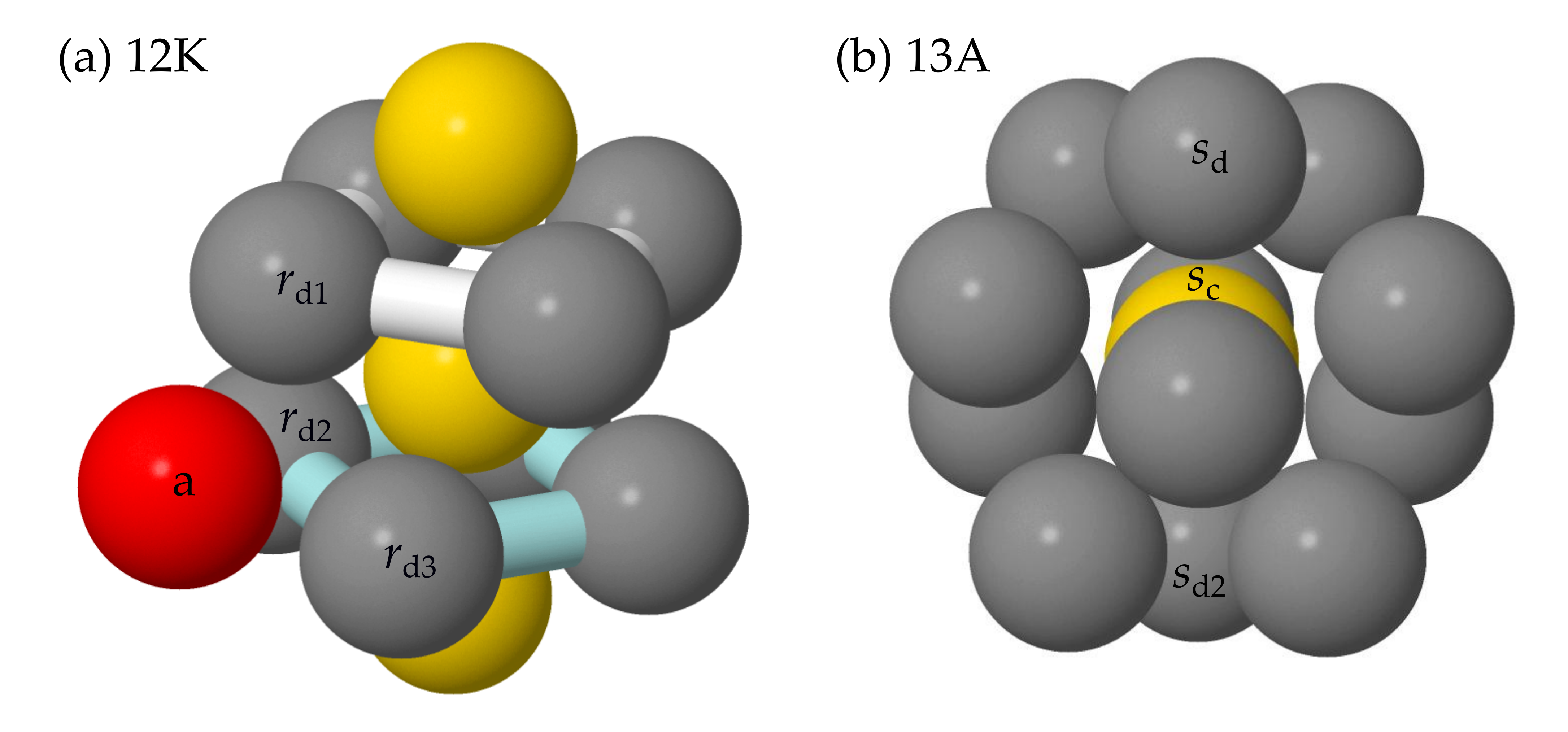} 
\par\end{centering}
\caption{The (a) 12K and (b) 13A clusters. (a) The additional particle $a$ is bonded to particles $r_{\mathrm{d}1}$, $r_{\mathrm{d}2}$ and $r_{\mathrm{d}3}$ in the sp4 rings of the 11A cluster. }
\label{fig12K_13A} 
\end{figure}

\textit{12K. --- } The 12K cluster is the minimum energy of KA Lennard-Jones mixture for twelve particles, where eight of the particles are the larger $A$-species [Fig.\ \ref{fig12K_13A}(a)]. See table \ref{tableDetection12K} for details.

\textit{13A. --- } The 13A cluster is the minimum energy for thirteen Morse particles with the range of the potential in $\rho_0 < 14.76$ [Fig.\ \ref{fig12K_13A}(b)]. This cluster is topologically equivalent to the icosahedral cluster discussed by Frank \cite{frank1952}. See table \ref{tableDetection12K} for details.

\begin{figure}
\begin{centering}
\includegraphics[width=7 cm]{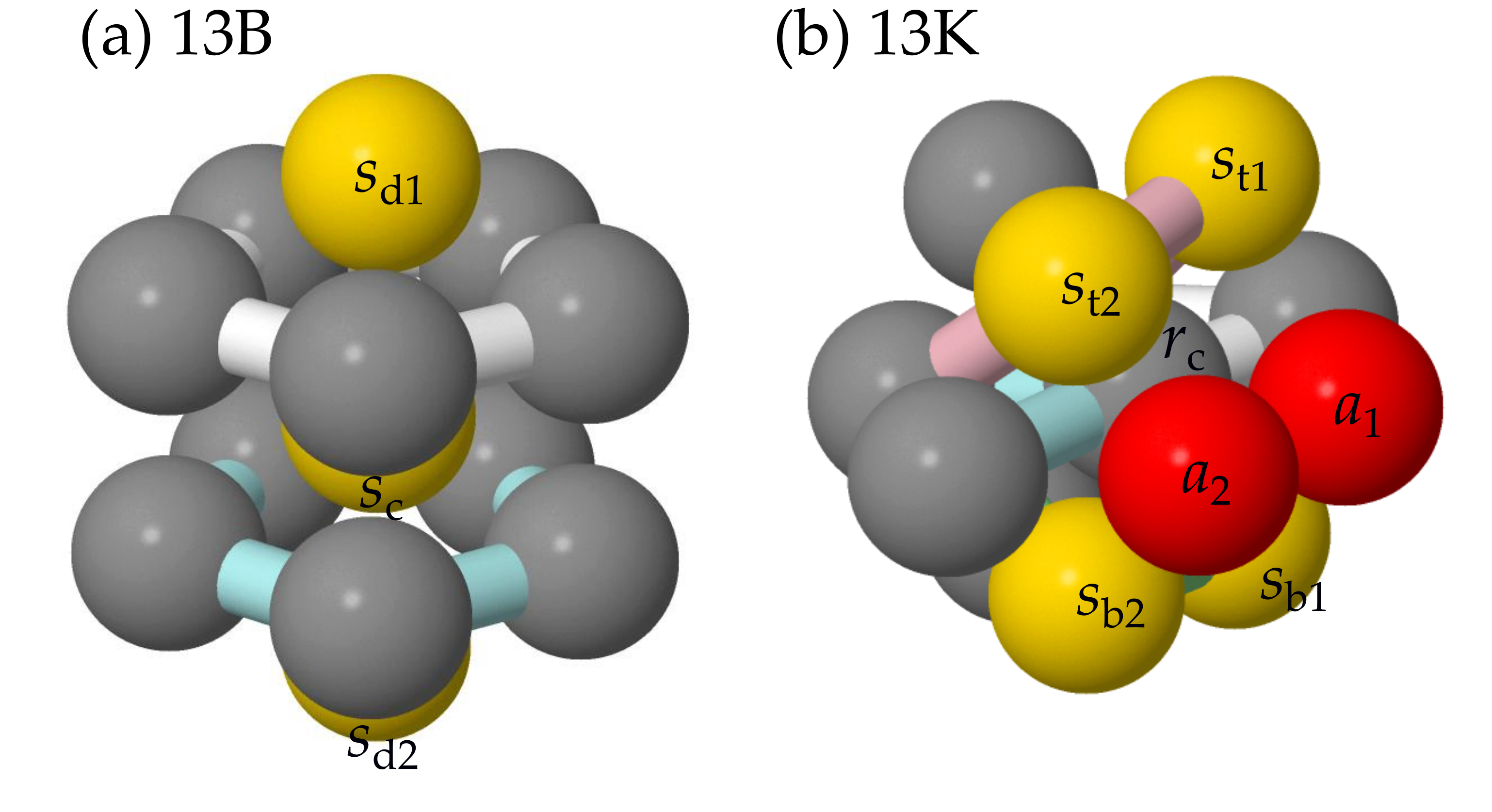} 
\par\end{centering}
\caption{The (a) 13B and (b) 13K clusters. (a) Each sp5 ring particle from one 7A clusters is bonded to a single sp5 ring particle in the other 7A cluster. (b) The sp3 rings of the additional 5A clusters are not shown.}
\label{fig13B_13K} 
\end{figure}

\textit{13B. --- } The 13B cluster is the minimum energy for thirteen Morse particles with the range parameter of the potential in $14.76\le\rho_0<25$  [Fig.\ \ref{fig13B_13K}(a)]. See table \ref{tableDetection12K} for details.

\textit{13K. --- } The 13K cluster is the minimum energy of KA Lennard-Jones mixture for thirteen particles, where seven of the particles are the larger $A$-species [Fig.\ \ref{fig13B_13K}(b)]. See table \ref{tableDetection12K} for details.

\section*{Crystal clusters}
\label{appendixCrystalClusters}

In order to detect crystalline structure and to test the performance of the TCC algorithm against phases with known structure, detection routines are implemented for crystalline clusters.

\begin{figure}
\begin{centering}
\includegraphics[width=7 cm]{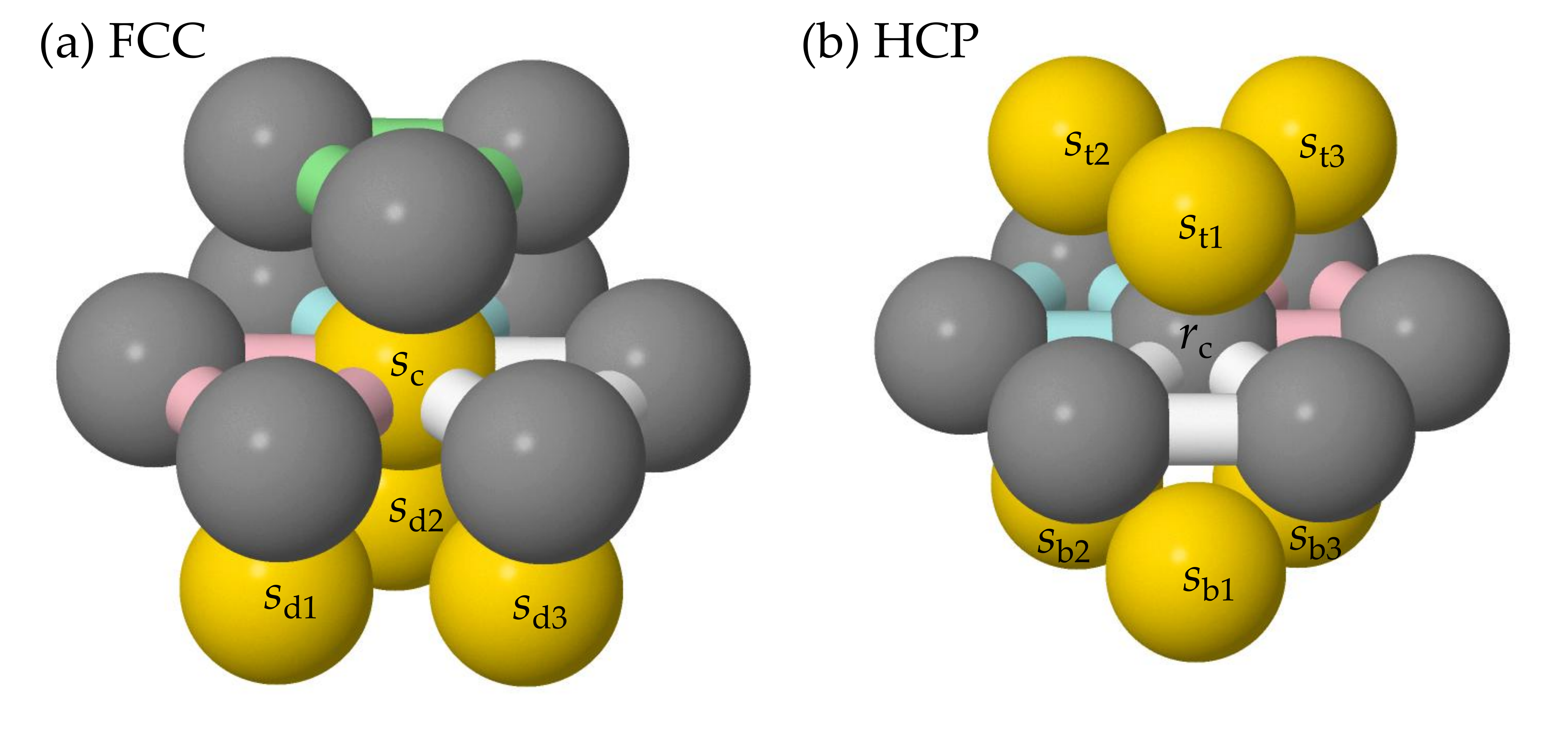} 
\par\end{centering}
\caption[The FCC and HCP clusters.]{The (a) FCC and (b) HCP clusters. (a) The six-membered ring consists of the gray particles in the horizontal plane around $s_\mathrm{c}$. (b) The six-membered ring consists of the gray particles in the horizontal plane around $r_\mathrm{c}$.}
\label{figFCC_HCP} 
\end{figure}

\textit{FCC. --- } The $m=13$ FCC cluster is a particle and its first-nearest neighbors as taken from a FCC crystal lattice [Fig.\ \ref{figFCC_HCP}(a)]. See table \ref{tableDetectionXtal} for details.

\textit{HCP. --- } The $m=13$ HCP cluster is a particle and its neighbors taken from a HCP crystal lattice [Fig.\ \ref{figFCC_HCP}(b)]. See table \ref{tableDetectionXtal} for details.

\begin{figure}
\begin{centering}
\includegraphics[width=3 cm]{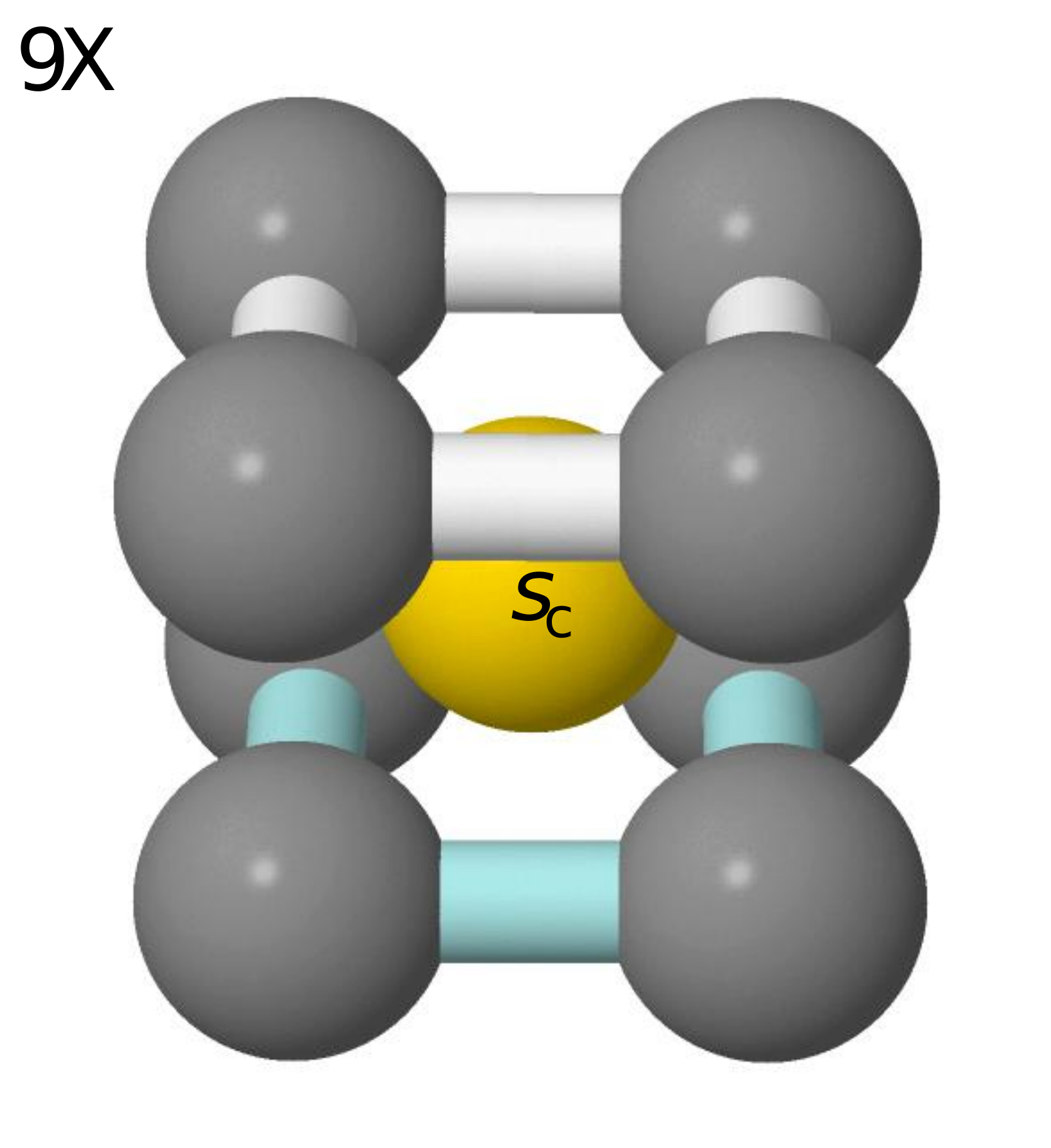} 
\par\end{centering}
\caption{The 9X cluster. Each particle in the sp4 rings is bonded to a single particle in the other sp4 ring.}
\label{fig9X} 
\end{figure}

\textit{9X. --- } The 9X cluster is found in the minimum energy FCC and BCC crystal lattices [Fig.\ \ref{fig9X}]. However the size of the cluster means that it does not uniquely determine crystalline order (in the same manner that tetrahedral order, although found in all bulk crystalline states, does not imply crystallinity). See table \ref{tableDetectionXtal} for details.

\begin{table*}															
\begin{centering}															
\begin{tabular}{lll}															
\hline														
\hline														
Cluster	&	Detection routine	&	Figure	\\
\hline																
6Z	&  A pair of 5A clusters where :	&   \ref{fig6Z_7K}(a)   \\
 	&  The are no common spindles between the two 5A clusters.	&     \\
	& One spindle of 5A$_i$ is common with a sp3 ring particle of 5A$_j$. & \\
	& One spindle of 5A$_j$ is common with a sp3 ring particle of 5A$_i$. & \\
	 & The spindles in the two sp3 rings are bonded. & \\
	& Two particles are common between the sp3 rings of 5A$_i$ and 5A$_j$. & \\  \hline
7K & A pair of 5A clusters where :	  & \ref{fig6Z_7K}(b)   \\
  	& 5A$_i$ and 5A$_j$ have one common spindle particle.  &   \\
	& The other spindle of 5A$_i$ is distinct from all the particles in 5A$_j$. &  \\
	& The other spindle of 5A$_j$ is distinct from all the particles in 5A$_i$.  &  \\
	& There are two common particles between the sp3 rings of 5A$_i$ and 5A$_j$.  &  \\  \hline
8A 	& A pair of sp5b clusters where:	  & \ref{fig8A_8B}(a)   \\
	&  The spindle particles are distinct. & \\
	& There are four common particles between sp5 rings of sp5b$_i$ and sp5b$_j$. & \\
\textbf{Or} & A pair of sp5b clusters where: & \\
	& Both 7A$_i$ spindle particles are common with the 7A$_j$ spindles. & \\
	& There are four common particles between sp5 rings of 7A$_i$ and 7A$_j$. & \\
\textbf{Or} & A sp5b cluster and a 7A cluster where: & \\
	& One 7A spindle is common with the sp5b spindle. & \\
	& The other 7A spindle is distinct from all the sp5b particles. & \\
	& There are four common particles between sp5 rings of sp5b and 7A. & \\  \hline
8B 	& A 7A cluster and with one additional particle where:	  & \ref{fig8A_8B}(b)   \\
	& The additional particle is bonded to one 7A spindle. & \\
	& The additional particle is bonded to two neighboring particles & \\
     	& in the sp5 ring particles of 7A. & \\ \hline
8K 	&  Three 5A clusters where: & \ref{fig8K_9A}(a) \\
	& Two common particles in the sp3 rings of 5A$_i$, 5A$_j$ and 5A$_k$. & \\
	& One common spindle between 5A$_i$ and 5A$_j$. & \\
	& One common spindle between 5A$_i$ and 5A$_k$. & \\
	& One common spindle between 5A$_j$ and 5A$_k$. & \\
	& One particle in sp3 ring of 5A$_i$ is not in 5A$_j$ and 5A$_k$. & \\
	& One particle in sp3 ring of 5A$_j$ is not in 5A$_i$ and 5A$_k$. & \\
	& One particle in sp3 ring of 5A$_k$ is not in 5A$_i$ and 5A$_j$. & \\ \hline
\hline														
\end{tabular}															
\end{centering}															
\caption{Detection routines for compound clusters for $6 \le m \le 8$.}
\label{tableDetection}
\end{table*}

\begin{table*}															
\begin{centering}															
\begin{tabular}{lll}															
\hline														
\hline														
Cluster	&	Detection routine	&	Figure	\\
\hline																
9A	&  Three sp4b clusters where :	&   \ref{fig8K_9A}(b)   \\
	& The three spindle particles are distinct. & \\
	& There are no bonds between the three spindle particles. & \\
	& There are two common particles between sp4 rings of sp4b$_i$ and sp4b$_j$. & \\
	& The distinct particles between the sp4 rings of sp4b$_i$ & \\
	& and sp4b$_j$ form the sp4 ring of sp4b$_k$. & \\ \hline 
9B	&  Three sp4b clusters where :	&   \ref{fig9B_9K}(a) \\
	& There is one common spindle particle. & \\
	& The distinct spindle particles are bonded. & \\
	& The distinct spindle of 7A$_i$ is common with an sp5 ring particle of 7A$_j$. & \\
	& The distinct spindle of 7A$_j$ is common with an sp5 ring particle of 7A$_i$. & \\
	& There are two common particles between the sp5 rings of 7A$_i$ and 7A$_j$. & \\ \hline
9K	&  A pair of 6A clusters where :	&   \ref{fig9B_9K}(b) \\
	& One common spindle particle. & \\
	& The uncommon spindle of 6A$_i$ is distinct from 6A$_j$. & \\ 
	& The uncommon spindle of 6A$_j$ is distinct from 6A$_i$. &	\\ 
	& There are two common particles between sp4 rings of 6A$_i$ and 6A$_j$. & \\ 
	& The common sp4 ring particles are bonded. & \\ \hline
10A & A pair of sp4b clusters where: &  \ref{fig10A_10B}(a) \\ 
	& All the particles in sp4b$_i$ and sp4b$_j$ are distinct. & \\
	& The spindle particles of sp4b$_i$ and sp4b$_j$ are not bonded. & \\
	& Each particle of sp4 ring of sp4b$_i$ is bonded to exactly two particles in sp4b$_j$. & \\
	& Each particle of sp4 ring of sp4b$_j$ is bonded to exactly two particles in sp4b$_i$. & \\ \hline
10B & A 9B and 7A cluster where: &  \ref{fig10A_10B}(b) \\ 
	& One spindle from 7A is common to the ``common spindle particle'' of the 9B cluster. & \\
	& The other spindle from 7A is bonded to the two distinct spindles of 9B. & \\
	& Two sp5 ring particles from 7A are common with the distinct spindles of 9B. & \\
	& Two sp5 ring particles from 7A are common with the distinct sp5 particles of 9B. & \\
	& The final sp5 ring particle from 7A is distinct from the 9B cluster. & \\ \hline
10K & A 9K cluster with an additional particle where: &  \ref{fig10K_10W}(a) \\ 
	& The common spindle particle in 9K has one additional neighbor that & \\
	& is distinct from all of the other 9K particles. & \\ \hline
10W & Six sp5b clusters where:: &  \ref{fig10K_10W}(b) \\ 
	& All of the spindles are common. & \\
	& The common spindle has coordination number 9. & \\ \hline 
\hline														
\end{tabular}															
\end{centering}															
\caption{Detection routines for compound clusters for $9 \le m \le 10$.}
\label{tableDetection9}
\end{table*}

\begin{table*}															
\begin{centering}															
\begin{tabular}{lll}															
\hline														
\hline														
Cluster	&	Detection routine	&	Figure	\\
\hline																
11A	& Two 6A clusters where:	&   \ref{fig11A_11B}(a)   \\
	& One common spindle particle. & \\
	& All other particles in 6A$_i$ are distinct from those in 6A$_j$. & \\
	& Each particle in the sp4 ring 6A$_i$ is bonded to two particles in sp4 ring of 6A$_j$. & \\
	& Each particle in the sp4 ring 6A$_j$ is bonded to two particles in sp4 ring of 6A$_i$. & \\  \hline
11B$^\dagger$	& A 9B cluster and two additional particles where$^\dagger$:	&   \ref{fig11A_11B}(a)   \\
	& The common spindle particle from the 9B cluster has coordination number 10. & \\
	& The two additional particles are bonded to each other and to the common spindle  & \\
	& particle of 9B. & \\
	& Each additional particle is bonded to two more particles in the shell of the & \\
	& 9B cluster, leading to a total of four bonds between the additional particles & \\
	& and 9B. & \\
	& For each additional particle, the two shell particles to which they are bonded & \\
	& are not themselves bonded. & \\
	& The four shell particles of the 9B cluster that are bonded to the two additional & \\
	& particles form two pairs that are neighbors. & \\ \hline
11C 	& Two 7A clusters where:	&   \ref{fig11C_11E}(a)   \\ 
	& The is one common spindle particle. \\
	& There are two further common particles between 7A$_i$ and 7A$_j$. \\
	& These are a bonded pair in the sp5 rings of 7A$_i$ and 7A$_j$. \\
	& There are two further bonds between distinct particles in the sp5 rings of 7A$_i$  & \\
	& and 7A$_j$. \\  \hline
11E 	& A 9B and 7A cluster where: & \ref{fig11C_11E}(b)   \\ 
	& One 7A spindle particle is common to one of the uncommon spindle & \\
	& particles, $s_{\mathrm{d}1}$, in the 7A clusters constituting the 9B cluster. & \\
	& The other spindle particle of the additional 7A is labeled $s_{\mathrm{d}3}$ and & \\
	& is bonded to the other uncommon spindle particle $s_{\mathrm{d}2}$ in 9B and & \\
	& the common spindle particle $s_\mathrm{c}$ of 9B. & \\
	& Of the 7A cluster sp5 ring particles, one is common to the common with & \\
	& $s_\mathrm{c}$, one is common with $s_{\mathrm{d}2}$, and one is common to & \\
	& one of the uncommon sp5 ring particles of the 9B cluster [$r$ in Fig.\ \ref{fig11C_11E}(b)]. & \\
	&The final two sp5 ring particles are distinct from the 9B cluster. & \\ \hline
\hline
\end{tabular}															
\end{centering}															
\caption{Detection routines for compound clusters for $11A$ to $11E$.
 $^\dagger$The particles in the 9B cluster that are not the common spindle are termed the \textit{shell} particles.	}
\label{tableDetection11}
\end{table*}

\begin{table*}															
\begin{centering}															
\begin{tabular}{lll}															
\hline														
\hline														
Cluster	&	Detection routine	&	Figure	\\
\hline																
11F	& Two 5A and two 6A clusters where:	&   \ref{fig11F_11W}(a)   \\
	& Each spindle in 5A$_i$ is bonded to one spindle in 5A$_j$ and vice versa. & \\
	& Thus there are two pairs of bonded spindles, ($s_{\mathrm{t}1}$,  $s_{\mathrm{t}2})$ and ($s_{\mathrm{b}1}$,$s_{\mathrm{b}2}$). & \\
	& There is one common particle between the sp3 ring of 5A$_i$ and 5A$_j$, $r_\mathrm{c}$. & \\
	& There is one bond between the other particles in sp3 rings of  & \\
	& 5A$_i$ and 5A$_j$, forming a pair ($r_{\mathrm{d}1}$,$r_{\mathrm{d}2}$). & \\
	& 6A$_k$ has one spindle in common with $r_\mathrm{c}$, & \\
	& and the other spindle is distinct from 5A$_i$, 5A$_j$ and 6A$_l$.  & \\
	& Its sp4 ring particles are $r_{\mathrm{d}1}$, $r_{\mathrm{d}2}$,  $s_{\mathrm{t}1}$, $s_{\mathrm{t}1}$. & \\
	& 6A$_l$ has one spindle in common with $r_\mathrm{c}$, and the other  & \\
	& spindle is distinct from 5A$_i$, 5A$_j$ and 6A$_k$.  & \\
	& Its sp4 ring particles are $r_{\mathrm{d}1}$, $r_{\mathrm{d}2}$, $s_{\mathrm{b}1}$, $s_{\mathrm{b}1}$. & \\ \hline
11W	& A 10B cluster and an additional particle where:	&   \ref{fig11F_11W}(b)   \\
	& The common spindle of the 10B cluster has coordination number 10. &  \\
	& The additional particle is not bonded to any of the distinct &  \\
	& spindles of the 7A clusters constituting the 10B cluster. &  \\ \hline
12A 	& An 11C cluster with an additional particle where:	& \ref{fig12A_12B}(a) \\ 
	& The common spindle particle of the 11C has coordination number 11. & \\ 
	& The additional particle is bonded to three particles in the 11C cluster: & \\ 
	& the common spindle, and the two sp5 ring particles of the 7A clusters & \\ 
	& constituting 11C that are not bonded to any of the other 7A cluster's  & \\ 
 	& particles [$r_{\mathrm{d}5}$ and $r_{\mathrm{d}6}$ in Fig.\ \ref{fig11C_11E}(a)]. & \\ \hline 
12B 	& 6 7A clusters where:	& \ref{fig12A_12B}(b) \\ 
	& There is one central 7A cluster with spindles $s_\mathrm{c}$ and s$_\mathrm{d}$. & \\ 
	& The other five 7A clusters have one spindle given by s$_\mathrm{c}$ and & \\ 
	& one spindle bonded to s$_\mathrm{d}$. & \\ \hline 
12D 	& A 7A cluster and an 11E cluster where: & \ref{fig12D_12E}(a) \\ 
	& The spindle particles of the 7A cluster are common with 11E cluster & \\ 
	& spindles $s_{\mathrm{d}2}$ and $s_{\mathrm{d}3}$. & \\ 
	& Of the sp5 ring particles of the 7A cluster, one is common to $s_\mathrm{c}$, & \\ 
	& one is common to $s_{\mathrm{d}1}$, two are in the sp5 rings of the 7A clusters & \\ 
	& constituting 11E [$r_{\mathrm{c}1}$ and $r_{\mathrm{c}2}$ in Fig.\ \ref{fig12D_12E}(a)], and one is new. & \\ \hline 
12E 	& An 11F cluster and a 5A cluster where: & \ref{fig12D_12E}(b) \\ 
	& The spindle atoms of the 5A cluster are common with the uncommon spindle & \\
	& atoms of the 6A clusters constituting the 11F cluster. & \\
	& Of the sp3 ring particles in the 5A cluster, two are common with $r_{\mathrm{d}1}$ & \\
	& and $r_{\mathrm{d}2}$ from the 11F cluster, and one is new. & \\ \hline
\hline
\end{tabular}															
\end{centering}															
\caption{Detection routines for compound clusters for $11F$ to $12E$.}
\label{tableDetection11F}
\end{table*}

\begin{table*}															
\begin{centering}															
\begin{tabular}{lll}															
\hline														
\hline														
Cluster	&	Detection routine	&	Figure	\\
\hline																
12K & An 11A cluster with one additional particle where:  & \ref{fig12K_13A}(a) \\ 
	& The additional particle is bonded to three mutually bonded sp4 ring & \\
	& particles in the 6A clusters that constitute the 11A cluster. & \\ \hline
13A 	&  An 12B cluster is supplemented with a 7A cluster where:  & \ref{fig12K_13A}(b) \\ 
	& The 7A cluster has one spindle given by $s_{\mathrm{c}}$ of the 12B cluster, and one & \\
	& spindle that is distinct from the 12B particles. & \\
	& The sp5 ring particles of the 7A cluster are distinct from the sp5 ring & \\
	& particles of the central 7A cluster in 12B. & \\ \hline
13B 	& Two 7A clusters where: &  \ref{fig13B_13K}(a) \\
	& There is one common particle between 7A$_i$ and 7A$_j$, which is a spindle $s_\mathrm{c}$. & \\
	& Other spindle particles are distinct and not bonded. & \\
	& Each particle from sp5 ring of 7A$_i$ is bonded to one sp5 particle of 7A$_j$. & \\
	& Each particle from sp5 ring of 7A$_j$ is bonded to one sp5 particle of 7A$_i$. & \\ \hline
13K 	& An 11F cluster and two 5A clusters where: &  \ref{fig13B_13K}(b)  \\
	& Cluster 5A$_i$ has spindles $s_{\mathrm{t}1}$, $s_{\mathrm{b}1}$. & \\
	& The sp3 ring particles of 5A$_i$ are $r_\mathrm{c}$ & \\
	& The other sp3 ring particle from the 5A cluster in 11F with spindles  & \\
	& $s_{\mathrm{t}1}$, $s_{\mathrm{b}1}$ that is not $r_{\mathrm{d}1}$ in Fig.\ \ref{fig11F_11W}(a). & \\
	& The third sp3 ring particle from 5A$_i$ is distinct from the 11F cluster. & \\
	& Cluster 5A$_j$ has spindles $s_{\mathrm{t}2}$, $s_{\mathrm{b}2}$. & \\
	& The sp3 ring particles of 5A$_j$ are $r_\mathrm{c}$ & \\
	& The other sp3 ring particle from the 5A cluster in 11F with spindles & \\
	& $s_{\mathrm{t}2}$, $s_{\mathrm{b}2}$ that is not $r_{\mathrm{d}2}$ in Fig.\ \ref{fig11F_11W}(a). & \\
	& The third sp3 ring particle from 5A$_j$ is distinct from the 11F cluster. & \\ \hline
\hline
\end{tabular}															
\end{centering}															
\caption{Detection routines for compound clusters for $12K$ to $13K$.}
\label{tableDetection12K}
\end{table*}

\begin{table*}															
\begin{centering}															
\begin{tabular}{lll}															
\hline														
\hline														
Cluster	&	Detection routine	&	Figure	\\
\hline																
FCC  &  Four sp3b or three sp3b and a 5A cluster where:  & \ref{figFCC_HCP}(a)  \\ 
	& The spindle particles of sp3b$_i$, sp3b$_j$, and sp3b$_k$ are all bonded to each other. & \\
	& There is one common particle $s_\mathrm{c}$ between sp3b$_i$, & \\
         & 3b$_j$, and sp3b$_k$, which is in the sp3 ring of each cluster. & \\
	& The rest of the particles are distinct. & \\
	& One of the uncommon sp3 ring particles in sp3b$_i$ is bonded to one uncommon & \\
         & 3 ring particle in sp3b$_j$, and the other is bonded to one uncommon ring particle  & \\
	& in sp3b$_k$.&  \\
	& One of the uncommon sp3 ring particles in sp3b$_j$ is bonded to one uncommon sp3 & \\
	& ring particle in sp3b$_i$, and the other is bonded to one uncommon  & \\
	& ring particle sp3b$_k$.  & \\
	& One of the uncommon sp3 ring particles in sp3b$_k$ is bonded to one uncommon sp3 & \\
	& ring particle in sp3b$_i$, and the other is bonded to one uncommon  & \\
	& ring particle in sp3b$_j$.  &\\
	& Excluding the spindle particles, we now have a six-membered ring of particles around & \\
	& $s_\mathrm{c}$. The six-membered ring defines six sp3 rings with $s_\mathrm{c}$, & \\
	& three of which were the sp3 rings for sp3b$_i$, sp3b$_j$, and sp3b$_k$, and three of  & \\
	& which are new. & \\
	& The fourth sp3b cluster or the 5A cluster have a spindle $s_\mathrm{c}$. & \\
	& Each sp3 particle in the fourth cluster forms an sp3b cluster with each of & \\
	&  the new sp3 & \\
	& rings from the six-membered ring around $s_\mathrm{c}$. & \\
	&  If the fourth cluster is a 5A cluster, its spindle that is not $s_\mathrm{c}$ is not counted & \\
	& as part of the FCC cluster. & \\ \hline
HCP & Three 5A clusters where:  & \ref{figFCC_HCP}(b)  \\ 
	& There is one common particle $r_\mathrm{c}$ between 5A$_i$, 5A$_j$ and 5A$_k$ that  & \\
	& is in the sp3 ring of each cluster. & \\
	& The spindle atoms from 5A$_i$, 5A$_j$ and 5A$_k$ form two sp3 rings ($s_{\mathrm{t}1}$, & \\ 
	& $s_{\mathrm{t}2}$, $s_{\mathrm{t}3}$) and ($s_{\mathrm{b}1}$, $s_{\mathrm{b}2}$, $s_{\mathrm{b}3}$). & \\
	& Within the three 5A clusters, the spindle atoms are only bonded to the particles from & \\
	& the cluster's own sp3 ring. & \\
	& The uncommon sp3 ring particles from 5A$_i$, 5A$_j$ and 5A$_k$ form a six-membered ring & \\
	& around $r_\mathrm{c}$, i.e.\ each is bonded to a single particle from any of the other cluster's & \\
	& uncommon sp3 ring particles. & \\ \hline
9X 	& Two sp4b or two 6A or an sp4b and a 6A cluster where:  & \ref{fig9X} \\ 
	& The is one common particle between the clusters, which is a spindle particle. & \\
	& Each of the particles in the two sp4 rings of the constituent clusters is bonded to one &  \\
	& particle in the other sp4 ring. &  \\   
	& NB.\ If identified with 6A clusters, the uncommon spindle atoms in 9X are not counted  & \\
	& as part of the cluster. &  \\  \hline
\hline
\end{tabular}															
\end{centering}															
\caption{Detection routines for crystalline clusters.}
\label{tableDetectionXtal}
\end{table*}

\end{document}